\documentclass[reprint,
superscriptaddress,
amsmath,
amssymb,
aps,
prb,
floatfix,
english
]{revtex4-2}

\usepackage{braket}
\usepackage{bibunits}
\usepackage[unicode=true, colorlinks=true, citecolor={blue!80!black}, urlcolor={blue!50!black}, linkcolor = {blue!80!black}]{hyperref}
\defaultbibliography{references}
\defaultbibliographystyle{apsrev4-2}
\usepackage{graphicx}
\usepackage{svg}
\usepackage{dcolumn}
\usepackage{bm}
\usepackage{siunitx}
\usepackage[T1]{fontenc}
\usepackage[utf8]{inputenc}
\usepackage{array}
\usepackage{multirow}

\begin{document}
\widetext

\title{On-demand population of Andreev levels by their ionization in the presence of Coulomb blockade}

\author{Pavel D.~Kurilovich}
\thanks{These two authors contributed equally.\\
pavel.kurilovich@yale.edu, vlad.kurilovich@yale.edu}
\affiliation{Department of Applied Physics, Yale University, New Haven, CT 06520, USA}
\affiliation{Department of Physics, Yale University, New Haven, CT 06520, USA}

\author{Vladislav D.~Kurilovich}
\thanks{These two authors contributed equally.\\
pavel.kurilovich@yale.edu, vlad.kurilovich@yale.edu}
\affiliation{Department of Physics, Yale University, New Haven, CT 06520, USA}

\author{Aleksandr E.~Svetogorov}
\affiliation{Fachbereich Physik, Universität Konstanz, D-78457 Konstanz, Germany}

\author{Wolfgang Belzig}
\affiliation{Fachbereich Physik, Universität Konstanz, D-78457 Konstanz, Germany}

\author{Michel H. Devoret}
\affiliation{Department of Applied Physics, Yale University, New Haven, CT 06520, USA}
\affiliation{Department of Physics, Yale University, New Haven, CT 06520, USA}

\author{Leonid I.~Glazman}
\affiliation{Department of Applied Physics, Yale University, New Haven, CT 06520, USA}
\affiliation{Department of Physics, Yale University, New Haven, CT 06520, USA}

\begin{abstract}
A mechanism to deterministically prepare a nanowire Josephson junction in an odd parity state is proposed. The mechanism involves population of two Andreev levels by a resonant microwave drive breaking a Cooper pair, and a subsequent ionization of one of the levels by the same drive.
Robust preparation of the odd state is allowed by a residual Coulomb repulsion in the junction.
A similar resonant process can also be used to prepare the junction in the even state.
Our theory explains a recent experiment [J. J. Wesdorp, et al., Phys.~Rev.~Lett.~{\bf 131}, 117001 (2023)].
\end{abstract}

\maketitle
\section{Introduction}
\label{sec:intro}

Andreev bound states are subgap supercurrent-carrying fermionic states localized in a weak link between superconducting leads. Every Andreev level accommodates different many-body configurations: it can be occupied by 0, 1 (with spin up or down), or 2 quasiparticles. Recent experiments with semiconducting nanowire Josephson junctions~\cite{Geresdi2017,Krogstrup2019} (as well as with atomic point contacts~\cite{Urbina2013,Urbina2015}) managed to reveal these different configurations by probing microwave responses of the weak link. This opens a pathway for using an Andreev level as a qubit. The even parity states form a basis for an Andreev pair qubit~\cite{Wendin2003}.
Another approach is to use as a qubit the spin of a single quasiparticle trapped on the level~\cite{Nazarov2003,Nazarov2010}. Both the Andreev spin and the Andreev pair qubits were realized experimentally~\cite{Urbina2015,Devoret2018,Devoret2020,Devoret2021}.

\begin{figure}[t]
  \begin{center}
    \includegraphics[scale=1]{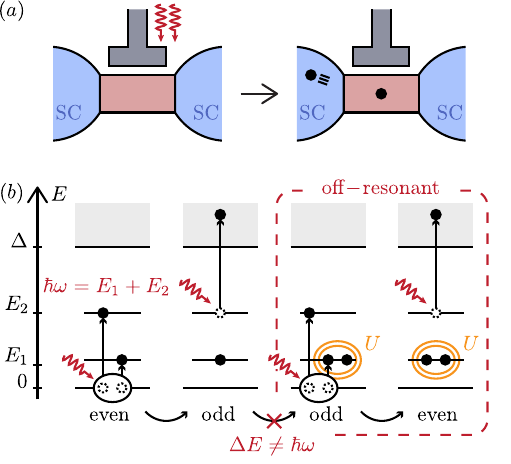}
    \caption{Deterministic population of an Andreev level with a \textit{single} quasiparticle. (a) A drive applied to the weak link breaks a Cooper pair leaving one quasiparticle in the Andreev level and ejecting another one to the superconducting leads. (b) Details of a resonant two-photon process preparing the weak link in the odd-parity state. 
    The link hosts two Andreev levels of energies $E_1$ and $E_2$, $E_2 > E_1$. The frequency of the drive tone is $\hbar\omega = E_1 + E_2$. Absorption of the first drive photon breaks a Cooper pair populating the two Andreev levels. Absorption of a second photon ejects the quasiparticle from the upper level into the leads ($\Delta$ is the gap in the leads). Further addition of quasiparticles is blocked by the Coulomb repulsion of strength $U$. As discussed in the text, a conceptually similar two-photon resonant process can be used to ionize the trapped quasiparticle bringing the weak link from the odd state into the even state.}
    \label{fig:summary}
  \end{center}
\end{figure}

To use Andreev levels as qubits, one needs to be able to initialize them in a given parity state. 
This task is simple for the application of the level as a pair qubit. Indeed, 
suppose the Andreev level  of energy $E_1$ is ``poisoned'' by an undesired quasiparticle (which may come, e.g., from the residual quasiparticle density in the leads~\cite{Devoret2004,Klapwijk2008,Echternach2008}).
Then, it is sufficient to apply a microwave tone of frequency $\hbar\omega > \Delta - E_1$ to ``evaporate'' this quasiparticle from the weak link into the leads ~\cite{Glazman2013,Pothier2013} (here $\Delta$ is the superconducting gap in the leads).

The situation is more complicated if the desired parity is odd. The na\"ive approach to the preparation of the odd state would be to irradiate the junction with microwaves of frequency $\hbar\omega > \Delta + E_1$. 
The latter condition allows the drive to break a Cooper pair putting one quasiparticle on the Andreev level and a second one into the continuum in the superconducting leads. This process brings the weak link to the odd parity sector. However, since the frequency of the drive also exceeds  {$(\Delta - E_1) / \hbar$}, the same drive will be capable of evaporating the quasiparticle from the level bringing the weak link back to the even parity state. 
The outlined initialization protocol thus unavoidably has a \textit{probabilistic} character.  Is it possible to \textit{deterministically} prepare the Andreev level in a state with a single quasiparticle?

Here, we answer this question affirmatively, and propose a mechanism by which an Andreev level can be deterministically prepared in a state with a single quasiparticle.  The mechanism relies on having at least two levels in the weak link (we denote their energies as $E_1$ and $E_2$){, and on Coulomb repulsion between quasiparticles populating the levels}. Its essence is summarized in Fig.~\ref{fig:summary}. Suppose a drive of frequency  {$(E_1 + E_2)/\hbar$} is applied to the junction. Such a resonant drive can break a Cooper pair populating each of the levels with a single quasiparticle. Under the right conditions on $E_{1,2}$ and $\Delta$, the same drive would then ionize the quasiparticle from the upper level, while leaving the quasiparticle in the lower level intact. An odd state would thus be prepared.

Crucially, preparation of the odd state relies on the residual Coulomb interaction in the weak link. Without the interaction, the drive would continue adding quasiparticles to the Andreev levels, and would continuously change  {the parity state} of the junction.
The Coulomb repulsion makes the process of the Cooper pair breaking off-resonant for the junction in the odd state, see the dashed box in Fig.~\ref{fig:summary}(b). Therefore, the quasiparticle addition automatically ceases once the junction reaches the desired state with a single quasiparticle.

 {We develop a phenomenological theory of the described 
mechanism of the odd-state preparation.
Our theory addresses how the inverse preparation time $\gamma$ depends on the power of the applied drive.
Additionally, we show that a resonant two-photon processes can also be used to clear the junction of quasiparticles, i.e., initialize the junction in the even state.}
As an illustration, we compute the parameters entering our phenomenological theory in a simple microscopic model, in which the weak link is treated as a quantum dot hosting two levels. 

Our theory provides an explanation for a recent experiment with a nanowire junction~\cite{jaap2021}. There, a strong microwave drive resonant with a transition in the even parity sector was found to change the junction parity from even to odd. This is surprising because the absorption of a microwave photon has to preserve the total fermion parity of the system.  {We explain this observation with our odd-state preparation mechanism}; the conservation of the total parity is ensured by an addition of an extra quasiparticle to one of the leads, see Fig.~\ref{fig:summary}(a). The measured dependence of the  {preparation rate} on the drive power is consistent with our results. Another, more restrictive mechanism relying on the presence of a hot photon bath was proposed in Ref.~\onlinecite{Yeyati2023}. Our approach 
avoids strong assumptions of Ref.~\onlinecite{Yeyati2023} about the electromagnetic environment of the junction, and sheds light on the unexplained trends observed in the experiment~\cite{jaap2021}.

\subsection*{Summary of results}
We consider a weak link hosting two Andreev levels with energies $E_1 < E_2 < \Delta$ [see Fig.~\ref{fig:summary}(b)].
For simplicity, we assume that both levels are spin-degenerate
(the role of the level splitting due to the spin-orbit coupling is discussed in Sec.~\ref{sec:concl}).
We label the many-body states of the system as $\ket{n_1, m_2}$, where $n, m \in \{0, 1, 2\}$ determine the number of quasiparticles in the first and the second levels, respectively \footnote{We dispense with the spin degree of freedom in the present discussion.}.
The ground state in the even and odd parity sectors are, respectively, $\ket{0_1, 0_2}$ and $\ket{1_1, 0_2}$.

 {Assume that the weak link is initially in the even-parity ground state, $\ket{0_1, 0_2}$.} Irradiation of the link by a continuous microwave tone at frequency $\omega = (E_1 + E_2) / \hbar$ {\it deterministically} brings it to  {the odd-parity} ground state [see Fig.~\ref{fig:summary}(b)].  {The mechanism works in the following way. While the link is still in the even state, the} drive coherently transfers population from $\ket{0_1, 0_2}$ to a state $\ket{1_1, 1_2}$. The system thus exhibits Rabi oscillations between these two states.
If the frequency of the drive is large enough,
\begin{equation}\label{eq:condition_1}
\hbar\omega = E_1 + E_2 > \Delta - E_2,
\end{equation}
then the same drive that induces the Rabi oscillations is also capable of evaporating the quasiparticle from the upper Andreev level. Therefore, over a sufficiently long time period, the quasiparticle gets ejected from the weak link; the system ends up in a state with a single quasiparticle in the lower level and another quasiparticle lost in the continuum.
The odd ground state $\ket{1_1, 0_2}$ is thus prepared.

\begin{figure}[t]
  \begin{center}
    \includegraphics[scale=1]{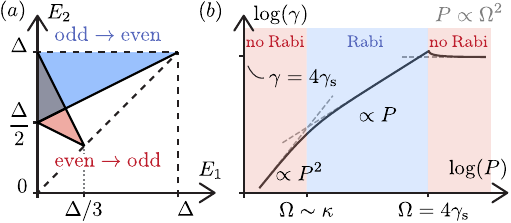}
    \caption{(a)  {Summary of conditions on the energies of the two Andreev levels under which the deterministic preparation of the odd-parity [Eqs.~\eqref{eq:condition_1} and \eqref{eq:condition_2}] and even-parity [Eq.~\eqref{eq:condition_3}] states are possible. (b) The dependence of the preparation rate of the odd-parity state $\gamma$ (i.e., the inverse time needed to prepare the state) on the drive power $P$. The drive is applied at resonance with the transition  $|0_1, 0_2\rangle \rightarrow |1_1, 1_2\rangle$, i.e.,} $\hbar\omega = E_1 + E_2$. When the matrix element of the drive  {$\Omega$} is small compared to the transition linewidth $\kappa$, the rate scales as $P^2$. At higher powers, the scaling is linear, $\gamma\propto P$. At the highest power, the rate saturates to a constant due to drive-induced broadening of the transition, see Eq.~\eqref{eq:sat_1}. The power dependence of the preparation rate of the even-parity state is qualitatively similar.}
    \label{fig:regimes}
  \end{center}
\end{figure}

In addition to inequality \eqref{eq:condition_1}, deterministic preparation of the odd state  {relies on} the fulfillment of two conditions.
 {The first one stems from the requirement that the drive should not be able to ionize the quasiparticle remaining in the lower Andreev level. This condition reads}
\begin{equation}\label{eq:condition_2}
\hbar\omega = E_1 + E_2 < \Delta - E_1.
\end{equation}
Conditions~\eqref{eq:condition_1} and \eqref{eq:condition_2} restrict the range of $E_1$ and $E_2$ for which the resonant quasiparticle generation is possible, as illustrated in Fig.~\ref{fig:regimes}(a). In what follows, we assume that conditions \eqref{eq:condition_1} and \eqref{eq:condition_2} are fulfilled.

 {Another requirement is that} residual Coulomb interaction has to be present in the weak link. Without interaction, the action of the drive does not ``switch off'' when the junction reaches the odd state. The drive continues to add and evaporate quasiparticles, and eventually brings the system back to the even parity state [the respective processes are illustrated in Fig.~\ref{fig:summary}(b) in a dashed box]. Coulomb interaction shifts the transition frequencies in the odd sector, making the drive off-resonant for a state with a single quasiparticle, $\ket{1_1,0_2}$. 
The odd ground state is thus stabilized. We assume that the Coulomb interaction is sufficiently weak such that---except for lifting the spectral degeneracy---its influence on energetics can be neglected [e.g., the influence on inequalities\eqref{eq:condition_1} and \eqref{eq:condition_2}].

The inverse time needed to prepare  {the odd-parity ground state} $\gamma$ depends on the power $P$ of the applied drive. To estimate $\gamma$, we note that the preparation is mediated by the interplay of the drive-induced transitions between $\ket{0_1, 0_2}$ and $\ket{1_1, 1_2}$, and the ionization of a quasiparticle from the second level. At a relatively low power of the drive, the former process is much quicker than the latter.
Indeed, the frequency of Rabi oscillations is determined by the drive matrix element $\Omega$ (which we measure in units of frequency),
\begin{equation}\label{eq:rabi}
\Omega_{\rm R} =  {2\Omega} \propto \sqrt{P}.
\end{equation}
By contrast, ionization of the second level happens with a rate that scales quadratically with $\Omega$,
\begin{equation}\label{eq:Gamma_i}
    \gamma_{\rm ion} = \frac{\Omega^2}{\gamma_{\rm s}} \propto P
\end{equation}
(the physical meaning of the frequency scale $\gamma_{\rm s}$ is the coupling between the higher ABS and the continuum due to the frive and will be clarified shortly).
For $\Omega \ll \gamma_{\rm s}$, the ionization occurs over many periods of Rabi oscillations. 
Because, on average, the system spends half of the time in the state $|1_1, 1_2\rangle$, we find the following expression for the rate $\gamma$ of the odd state preparation:
\begin{equation}
    \label{eq:half_of_ion}
       \gamma = \frac{1}{2} \gamma_{\rm ion} \propto P, 
\end{equation}
where $\gamma_{\rm ion}$ is given by Eq.~\eqref{eq:Gamma_i}.

Equations \eqref{eq:Gamma_i} and \eqref{eq:half_of_ion} show that the rate of the odd state preparation increases with power. This trend, however, breaks down at $\Omega \sim \gamma_{\rm s}$ when the Rabi frequency becomes comparable to $\gamma_{\rm ion}$. In fact, at higher powers, the rate ceases to change and saturates at $\gamma \sim \gamma_{\rm s}$. 
To understand this saturation, note that ionization transitions broaden the second level effectively spreading it over energy interval of width $\gamma_{\rm ion}$, cf.~Eq.~\eqref{eq:Gamma_i}. This broadening results in density of final states $\sim 1/\gamma_{\rm ion}$ for the transition from $\ket{0_1, 0_2}$ under the action of the drive. The corresponding transition rate can be calculated by Fermi's Golden rule, \begin{equation}\label{eq:sat_1}
        \gamma = 2\pi \Omega^2 \cdot \frac{1}{\pi \gamma_{\rm ion} / 2} = 4\gamma_{\rm s} \propto {\rm const}.
\end{equation}
Here, we used $\Omega$ for the transition matrix element in the first equality and Eq.~\eqref{eq:Gamma_i} for $\gamma_\mathrm{ion}$. A detailed derivation of Eq.~\eqref{eq:sat_1} is presented in Sec.~\ref{sec:res_gen}.

In the above, we disregarded the possibility of the quasiparticle recombination, $|1_1, 1_2\rangle \rightarrow |0_1, 0_2\rangle$. Recombination strongly affects the low-power tail of the dependence of $\gamma$ on $P$; it renders $\gamma \propto P^2$ when $\Omega \lesssim \kappa$, where $\kappa$ is the recombination rate.
This can be shown in the following way. To start with, we note that at $\Omega \lesssim \kappa$, one can estimate the population $p_{11}$ of $|1_1, 1_2\rangle$ from the generation-recombination balance. Recombination happens with a rate $\kappa$, whereas the generation rate can be estimated as $\Omega^2 / \kappa$. Therefore, $p_{11} \sim \Omega^2 / \kappa^2$. 
The odd parity state is reached if the generation process of a pair $|1_1, 1_2\rangle$ is followed by the ionization of a quasiparticle from the second level. Because the latter process happens with a rate $\gamma_{\rm ion}$, we can estimate the overall rate of the odd state preparation as $\gamma \sim \gamma_{\rm ion} p_{11}$. Combing the expression for $p_{11}$ with Eq.~\eqref{eq:Gamma_i}, we find 
\begin{equation}\label{eq:low_power}
        \gamma  =  \frac{ {4}\Omega^4}{\gamma_{\rm s} \kappa^2}  \propto P^2.
    \end{equation}
{The specific numeric coefficient follows from the detailed calculation described in Sec.~\ref{sec:res-P2-to-P}.}

The power dependence of the odd state preparation rate is summarized in Fig.~\ref{fig:regimes}(b).
In Ref.~\onlinecite{jaap2021}, the fit of the observed power-dependence of the odd state preparation rate by $\gamma \propto P^\alpha$ {(performed in a limited interval of $P$)} resulted in $\alpha = 1.5$. This falls between the predictions of Eqs.~\eqref{eq:half_of_ion} and \eqref{eq:low_power}.

A similar resonant two-photon process can also be used to ionize a quasiparticle from an Andreev level, i.e., to prepare the system in the even ground state.
This requires driving the junction at frequency  {$\omega = (E_2 - E_1) / \hbar$}.
Absorption of the first photon excites the quasiparticle from level $E_1$ to level $E_2$. Then, absorption of the second photon promotes the quasiparticle from $E_2$ into the continuum. 
The latter process is possible only if
\begin{equation}
\label{eq:condition_3}
    \hbar\omega = E_2 - E_1 > \Delta - E_2
\end{equation}
 [see Fig.~\ref{fig:regimes}(a)].
The energetic constraints for preparation of the even state are more loose than those for the preparation of the odd state. In particular, deterministic ionization is possible even in the absence of Coulomb interaction.
We note that the dependence of even state preparation rate $\tilde{\gamma}$ on the drive power is qualitatively similar to that of the odd state preparation. It crosses over from $\tilde{\gamma} \propto P^2$ at low powers, to $\tilde{\gamma} \propto P$ at intermediate powers, to saturation at a certain value $\tilde{\gamma} = 4\tilde{\gamma}_\mathrm{s}$ at high powers.

Finally, let us comment on the role of non-equilibrium quasiparticles ubiquitous in superconducting devices~\cite{Kaplan1976,Devoret1994,Larkin2006}. 
Such quasiparticles stochastically ``poison'' the weak link, and the parity of the latter randomly changes between being even and odd in the absence of the microwave drive.
 {In Sec.~\ref{sec:qps},} we show that even in the presence of stochastic switching, the desired parity state can be prepared with a high fidelity by applying a sufficiently strong drive.

Interestingly though, 
there is a limit on how close the fidelity can approach unity in the high-power regime.
For concreteness, let us focus on the preparation of the even state. In this case, the maximal attainable fidelity is limited by the saturation of the rate with the increase of power [see discussion after Eq.~\eqref{eq:condition_3}]. Accordingly, for the maximal probability of the even state $p_\mathrm{e}^\mathrm{max}$ we get
\begin{equation}
\label{eq:even_max}
    1 - p_\mathrm{e}^\mathrm{max} \sim   \frac{\Gamma_{\rm eo}}{\tilde{\gamma}_{\rm s}},
\end{equation}
where $\Gamma_{\rm eo}$ is the rate of transitions from even to odd parity due to the quasiparticle poisoning rate. The counterpart of Eq.~\eqref{eq:even_max} to the odd-state preparation includes the additional restriction coming from requirements on the strength of Coulomb interaction, cf.~Eq.~\eqref{eq:po-vs-U}.

In Sec.~\ref{sec:micro}, we illustrate our phenomenological theory by considering a concrete microscopic model of the weak link. We evaluate the phenomenological parameters and show how they depend on the phase bias across the link.
 
\section{Phenomenological model}
\label{sec:phenom}
In this section, we present a phenomenological theory of deterministic odd-state preparation. Within its framework, we derive the dependence of the inverse preparation time on the power of the applied resonant drive, as well as on the detuning of the drive from resonance. 
The developed theory is also suitable to describe a similar process used to prepare the even state, as we show in Section~\ref{sec:res-ion}.

\subsection{Preparation of the odd state\label{sec:res_gen}}
To describe the mechanism of the odd state preparation, we assume that the weak link hosts two Andreev levels with energies $E_1, E_2 < \Delta$. We consider the system initialized in an even state $\ket{0_1, 0_2}$. To populate the lowest level with a quasiparticle, one applies a drive at the frequency $f = (E_1 + E_2) / h$, as explained in introduction [also see Fig.~\ref{fig:summary}(b)]. The drive couples the state $\ket{0_1, 0_2}$ to an excited state $\ket{1_1, 1_2}$. It further couples the excited state to a continuum of states, in which one quasiparticle occupies the lower Andreev level and the second quasiparticle has energy above the superconducting gap. We denote such states as $\ket{1_1, 0_2, 1_c}$, where $c$ label the states with $E > \Delta$. We assume that the quasiparticle with above-the-gap energy never returns to the weak link. In this sense, state $\ket{1_1, 0_2, 1_c}$ describes the weak link in the odd ground state. The outlined mechanism works only if conditions \eqref{eq:condition_1} and \eqref{eq:condition_2} are fulfilled, and also there is a residual Coulomb interaction in the weak link (see discussion below). We assume that all of these requirements are satisfied.

We describe the system with a Hamiltonian
\begin{equation}\label{eq:H}
    H = H_0 + H_{\rm drive}(t).
\end{equation}
Here $H_0$ is the static part, in which we account for two discrete many-body states $\ket{0_1, 0_2}$ and $\ket{1_1, 1_2}$, as well as continuum of the high-energy excitations $\ket{1_1, 0_2, 1_c}$:
\begin{align}
    H_0 = (E_1 + E_2) \ket{1_1, 1_2}\bra{1_1, 1_2} + &\notag\\
    +\sum_{c} (E_1 + E_c)\ket{1_1, 0_2, 1_c}&\bra{1_1, 0_2, 1_c},
\end{align}
where $c$ labels single-particle states with $E_c > \Delta$; we take the energy of the state $\ket{0_1, 0_2}$ to be zero. Term $H_{\rm drive}(t)$ in Eq.~\eqref{eq:H} describes the drive applied to the junction. We consider $H_{\rm drive}(t)$ of the form
\begin{align}
    H_{\rm drive}(t) = \hbar\Omega e^{-i\omega t} \ket{1_1,1_2}\bra{0_1,0_2} &+ \notag\\+\hbar\Omega e^{-i\omega t} \sum_c \alpha_c \ket{1_1,0_2,1_c}&\bra{1_1,1_2} + \mathrm{h.c.},\label{eq:Vt}
\end{align}
where $\Omega$ is the drive amplitude.
Dimensionless numbers $\alpha_c$ characterize the strength of coupling to the continuum states.
In the above, we assume that the frequency of the drive $\omega$ is close to the transition frequency between the discrete states,  {$\omega \sim (E_1 + E_2)/\hbar$}.
Because of that, we include only the resonant terms in $H_{\rm drive}(t)$, 
{and dispense with the off-resonant ones such as  $e^{i\omega t}\ket{1_1, 1_2}\bra{0_1,0_2}$}. This is  {justified provided $\hbar|\Omega| \ll E_1 + E_2$}. 

{In Eq.~\eqref{eq:Vt}, we also neglected terms describing transition $\ket{1_{1}, 0_{2}, 1_c} \rightarrow \ket{2_{1}, 1_{2}, 1_c}$. This is justified only in the presence of a sufficiently strong Coulomb interaction. Indeed, without the interaction, the frequency of the latter transition,  {$\omega = (E_1 + E_2) / \hbar$}, would coincide with that of $\ket{0_1, 0_2} \rightarrow \ket{1_1, 1_2}$. Therefore, the drive applied at this frequency would be capable not only of changing parity from even to odd, but also of reversing it back [see Fig.~\ref{fig:summary}(b)]. Coulomb repulsion offsets the frequency of the transition $\ket{1_{1}, 0_{2}, 1_c} \rightarrow \ket{2_{1}, 1_{2}, 1_c}$ to  {$(E_1 + E_2 + U)/\hbar$}, where $U$ is the strength of repulsion. Therefore, the latter transition becomes off-resonant with the drive applied at $\omega = (E_1 + E_2) / \hbar$. Then, as long as the drive power is not too strong,  {$|\Omega| \ll U / \hbar$}, the odd-to-even parity switching does not occur.}

To compute the rate of the odd-state preparation, we consider the system initialized in the even-parity ground state, $|0_1,0_2\rangle$, and solve the time-dependent Schr\"odinger equation for the Hamiltonian \eqref{eq:H}. In this way, we obtain the probability to find the system in the odd state $w(t)$ as a function of time. To begin with, we parametrize the wavefunction as
\begin{align}
    &|\Psi(t)\rangle = \Psi_{00}(t)e^{i\omega t} \ket{0_1, 0_2} + \notag \\
    & + \Psi_{11}(t) \ket{1_1, 1_1} + \sum_c\Psi_{c}(t) e^{-i\omega t}\ket{1_1, 0_2, 1_c},
\end{align}
where $\Psi_{00} (0) = 1$, $\Psi_{11}(0) = \Psi_c(0) = 0$ at $t = 0$.
The desired probability to find the system in the odd state at time $t$ is given by $w(t) = \sum_c|\Psi_c(t)|^2$. Practically though, it is more convenient to express $w(t)$ as
\begin{equation}\label{eq:unitarity}
w(t) = 1 - |\Psi_{00}(t)|^2 - |\Psi_{11}(t)|^2
\end{equation}
{(whose equivalence to the initial definition follows from the probability conservation)}, and then focus on finding probabilities $|\Psi_{00}(t)|^2$ and $|\Psi_{11}(t)|^2$.

{The time-dependent Schr\"odinger equation for $|\Psi(t)\rangle$ results in the following system of equations for amplitudes $\Psi_i(t)$:}
\begin{subequations}
\label{eq:system_psi}
{\begin{align}
    i\dot{\Psi}_{00} &= \omega \Psi_{00} + \Omega^\star \Psi_{11} + i\delta(t),\label{eq:psi1}\\
    i\dot{\Psi}_{11} &= \frac{E_1 + E_2}{\hbar} \Psi_{11} + \Omega \Psi_{00} + \Omega^\star\sum_{c} \alpha^\star_c\Psi_c,\\
    i\dot{\Psi}_c &= \left(\frac{E_1 + E_c}{\hbar} - \omega \right) \Psi_c + \Omega \alpha_c\star\Psi_{11}\label{eq:psim}.
\end{align}}
\end{subequations}
The delta-function accounts for the initial conditions. In the frequency domain, system~\eqref{eq:system_psi} reduces to a system of algebraic equations which can be readily solved. By finding the solution and converting it back into the time domain, we obtain
\begin{align}\label{eq:psi1-sol}
{\Psi}_{00}(t) = \int \frac{d{\omega}^\prime}{2\pi}\,\frac{i  e^{-i{\omega}^\prime t}}{{\omega}^\prime + i0 - \omega  - \Sigma_{\Omega}({\omega}^\prime)/\hbar},
\end{align}
where
\begin{equation}\label{eq:Sigma_Omega}
    \frac{\Sigma_{\Omega}({\omega}^\prime)}{\hbar} = \frac{|\Omega|^2}{{\omega}^\prime + i0 - (E_1 + E_2)/\hbar - \Sigma_{c}({\omega}^\prime)/\hbar}
\end{equation}
and
\begin{equation}\label{eq:sigmac}
    \frac{\Sigma_{c}({\omega}^\prime)}{\hbar} = \sum_{c} \frac{|\Omega|^2|\alpha_c|^2}{{\omega}^\prime + i0 - (E_1 + E_c)/\hbar + \omega}.
\end{equation}
Self-energy functions $\Sigma_{\Omega}(\omega)$ and $\Sigma_{c}(\omega^\prime)$ describe, respectively, the drive-mediated coupling between states $\ket{0_1, 0_2}$ and $\ket{1_1, 1_2}$, and the coupling of $\ket{1_1, 1_2}$ to the continuum. Expression for $\Psi_{11}(t)$ can be found in a similar way.

Computing the integral in Eq.~\eqref{eq:psi1-sol} requires the knowledge of the self-energy function $\Sigma_{\Omega}(\omega^\prime)$. To evaluate this function, in Eq.~\eqref{eq:Sigma_Omega}, we neglect the variation of $\Sigma_{c}(\omega^\prime)$ with $\omega^\prime$, and treat $\Sigma_{c}(\omega^\prime)$ as a constant (complex) number. This approximation is justified as long as $\hbar\Omega \ll \Delta$. Indeed, $\Sigma_{c}(\omega^\prime)$ varies with $\omega^\prime$ on a scale set by the superconducting gap in the leads $\Delta$. At the same time, the integral in Eq.~\eqref{eq:psi1-sol} is determined by frequencies $\omega^\prime$ in a narrow vicinity $\sim \hbar\Omega$ of $\omega^\prime = (E_1 + E_2)/\hbar$. This allows us to change $\Sigma_{c}(\omega^\prime) \rightarrow \Sigma_{c}([E_1 + E_2]/\hbar)$ in Eq.~\eqref{eq:Sigma_Omega}. In the following, we suppress the argument and use a shortened notation $\Sigma_{c} \equiv \Sigma_{c}([E_1 + E_2]/\hbar)$.

To proceed, we decompose $\Sigma_{c}$ in its real and imaginary parts:
\begin{equation}\label{eq:SigmaReIm}
    \Sigma_{c} = {\rm Re}\,\Sigma_{c} + i\,{\rm Im}\,\Sigma_{c}.
\end{equation}
The real part describes the ac-Stark shift of the second level under the influence of the drive (see Appendix~\ref{sec:stark} for a detailed discussion). The imaginary part of $\Sigma_c$ describes {the ionization processes, i.e., the }transitions from $\ket{1_1, 1_2}$ into the odd state (with an extra quasiparticle in the continuum).
We express the imaginary part of the self-energy as
\begin{align}\label{eq:ImSigma}
    \frac{{\rm Im}\,\Sigma_{\rm c}}{\hbar} = -\pi|\Omega|^2 \sum_{c} |\alpha_c|^2 \delta \Bigl(\omega  - \frac{E_c - E_2}{\hbar}\Bigr) = - \frac{|\Omega|^2}{2\gamma_{\rm s}}.
\end{align}
Parameter $\gamma_{\rm s}$ here is determined by the density of states in the continuum $\nu(E_2 + \omega)$, as well as by the coupling $\alpha_c$  {of the upper Andreev} level to these states:
\begin{equation}\label{eq:gamma_s_matrix_el}
\gamma_{\rm s}^{-1} = \hbar
\,\nu(E_2 + \hbar\omega) |\alpha_c|^2.
\end{equation}
The physical significance of $\gamma_{\rm s}$ will become apparent shortly [see the discussion around Eqs.~\eqref{eq:prep_under} and \eqref{eq:prep_over} and Fig.~\ref{fig:regimes}(b)].
The ionization rate is given by
\begin{equation}
    \label{eq:ioniz}
    \gamma_{\rm ion} = -2\,{\rm Im}\,\Sigma_{c}/\hbar = \frac{|\Omega|^2}{\gamma_{\rm s}}.
\end{equation}
It scales linearly with the drive power $P$, $\gamma_{\rm ion} \propto |\Omega|^2\propto P$.

 {We now use the derived equations to obtain the probability $w(t)$ of finding the system in the odd state in the case of the drive applied at resonance, $\hbar\omega = E_1 + E_2 + {\rm Re}\,\Sigma_{c}$ [the last term accounts for the ac Stark shift, see Appendix~\ref{sec:stark} for details].
Computing the integral in Eq.~\eqref{eq:psi1-sol}, we find
\begin{align}
\label{eq:time-dyn}
|\Psi_{00}(t)|^2 = e^{-t \frac{|\Omega|^2}{2\gamma_{\rm s}}}\Bigl|
\frac{s + i\frac{|\Omega|^2}{4\gamma_{\rm s}}}{2s} e^{- i s t} +
(s \rightarrow -s)
\Bigr|^2,
\end{align}
where
\begin{equation}
    s \equiv s(\Omega) = \sqrt{|\Omega|^2 - \frac{|\Omega|^4}{16\gamma_{\rm s}^2}}.
\end{equation}
A similar calculation for $|\Psi_{11}|^2$ yields
\begin{equation}
    |\Psi_{11}|^2 = e^{-t\frac{|\Omega|^2}{2\gamma_{\rm s}}} |\Omega|^2 \sin^2 (st) / s^2. 
\end{equation}
The use of the above results in Eq.~\eqref{eq:unitarity} shows that the probability $w(t)$ of finding the system in the odd state grows monotonically with time, approaching unity at $t \rightarrow +\infty$. We can represent the result as $1 - w(t) = \sum_{i} c_i \exp(- \lambda_i t)$, where ${\rm Re}\,{\lambda_i} > 0$ for all $i$. We define the rate of the preparation of the odd state $\gamma$ as the smallest decrement ${\rm Re}\,\lambda_i$. We find for the rate:
\begin{equation}\label{eq:prep_rate_final}
    \gamma = {\rm Re}\,\Bigl[\frac{|\Omega|^2}{2\gamma_{\rm s}} - \sqrt{\frac{|\Omega|^4 }{4\gamma_{\rm s}^2} - 4|\Omega|^2}\Bigr].
\end{equation}}

If the amplitude of the drive is small, $|\Omega| < 4 \gamma_{\rm s}$, then the second term under the bracket in Eq.~\eqref{eq:prep_rate_final} is purely imaginary and thus drops out. As a result, we obtain
\begin{equation}\label{eq:prep_under}
    \gamma(|\Omega| < 4\gamma_{\rm s}) = \frac{|\Omega|^2}{2 \gamma_{\rm s}}.
\end{equation}
Since $|\Omega|^2 \propto P$, the odd-state preparation rate scales linearly with the power of the drive. 

Notice that, up to a factor of $1 / 2$, $\gamma$ coincides with the ionization rate of the upper Andreev level, cf.~Eqs.~\eqref{eq:ioniz} and \eqref{eq:prep_under}.
This feature reflects the character of the dynamics at low drive amplitudes, $|\Omega| < 4 \gamma_{\rm s}$. The latter condition defines the ``underdamped'' regime of the dynamics; in it, the system undergoes Rabi oscillations between states $\ket{0_1, 0_2}$ and $\ket{1_1, 1_2}$, see Eq.~\eqref{eq:time-dyn}. In the course of oscillations, the system spends a half of the time in a state $\ket{1_1, 1_2}$, which can be ionized by a drive. This is why $\gamma = \gamma_{\rm ion} / 2$.

Increasing the drive amplitude above $|\Omega| = 4\gamma_{\rm s}$ brings the dynamics into an ``overdamped'' regime, which is characterized by the absence of Rabi oscillations. In this regime, the second term in the square brackets of Eq.~\eqref{eq:prep_rate_final} is real leading to
\begin{equation}\label{eq:prep_over}
    \gamma(|\Omega| > 4\gamma_{\rm s}) = \frac{|\Omega|^2}{2 \gamma_{\rm s}} - \sqrt{\frac{|\Omega|^4 }{4\gamma_{\rm s}^2} - 4|\Omega|^2}.
\end{equation}
Equations \eqref{eq:prep_under} and \eqref{eq:prep_over} show that dependence of $\gamma$ on $|\Omega|$ is non-analytic, with a cusp at $|\Omega| = 4\gamma_{\rm s}$. In fact, the preparation rate reaches its maximum value at the cusp, $\gamma_{\rm max} = 8\gamma_{\rm s}$. Further increase of the drive amplitude leads to a  {gradual decrease} in $\gamma$.
Interestingly, 
the preparation rate saturates at a power-independent value $\gamma = 4\gamma_{\rm s}$ in the limit of high drive power, $|\Omega| \gg \gamma_{\rm s}$ [see the discussion around Eq.~\eqref{eq:sat_1} for the qualitative explanation of the saturation].

We note that for typical experimental parameters $\gamma_{\rm s} / 2\pi$ is at least of the order of a few GHz (see Section~\ref{sec:micro}). Achieving the Rabi rate $|\Omega| \sim \gamma_{\rm s}$ would most likely require drive power higher than that accessible in experiments. Therefore, practically, $\gamma_{\rm s}$ should not be a limiting factor for the odd-state preparation rate $\gamma$.

\subsection{Effects of quasiparticle recombination in the resonant case\label{sec:res-P2-to-P}}
In the previous section, we neglected the quasiparticle recombination, i.e., relaxation from $\ket{1_1, 1_2}$ to $\ket{0_1, 0_2}$. This led to the conclusion that the rate of the odd state preparation scales as $\propto P$ at the lowest drive powers $P$. However, the linear trend breaks down in the presence of recombination.
If the recombination rate $\kappa$ exceeds the drive amplitude, $|\Omega| \lesssim \kappa$, then no Rabi oscillations develop; the population transfer from $\ket{0_1, 0_2}$ to $\ket{1_1, 1_2}$ becomes suppressed. This results in the suppression of the odd-state preparation rate at low powers to $\propto P^2$, as we now show \footnote{For simplicity, we neglect an alternative decay channel by which state $\ket{1_1, 1_2}$ can relax to $\ket{0_1, 0_2}$. In this channel, the quasiparticle in the upper level first relaxes to the lower level, $\ket{1_1, 1_2}\rightarrow \ket{2_1, 0_2}$, and then the two quasiparticles in the lower level recombine, $\ket{2_1, 0_2}\rightarrow\ket{0_1, 0_2}$. Taking this decay channel into the account does not qualitatively alter our results.}.

To begin with, we neglect the ionization of the upper Andreev level. In that case, the dynamics of the system can be described with a two-level Bloch equation for the density matrix. According to this equation, the system reaches a steady state in which the probabilities of the states $|0_1,0_2\rangle$ and $|1_1,1_2\rangle$---$p_{00}$ and $p_{11}$, respectively---are related by
\begin{equation}
    \label{eq:11-vs-00}
    p_{11} = \frac{|\Omega|^2}{|\Omega|^2 + \kappa^2 / 4}p_{00}.
\end{equation}
In the absence of ionization, the total probability of states $|0_1,0_2\rangle$ and $|1_1,1_2\rangle$ is conserved, $p_{00} + p_{11} = 1$. Ionization leads to the decay of the probability which signifies the transition to the odd state. Since ionization is only possible from state $|1_1,1_2\rangle$, the decay can be described with the rate equation
\begin{equation}
    \label{eq:prob-decay}
    \frac{d}{dt}(p_{00} + p_{11}) = - \gamma_{\rm ion} p_{11},
\end{equation}
where $\gamma_{\rm ion}$ is given by Eq.~\eqref{eq:ioniz}. The probability to find the system in the odd state at time $t$ can be expressed as $w(t) = 1  - p_{00}(t) - p_{11}(t)$. Combining Eqs.~\eqref{eq:11-vs-00} and \eqref{eq:prob-decay} we find $w(t) = 1 - e^{-\gamma t}$. Here the inverse odd-state preparation time $\gamma$ is given by
\begin{equation}\label{eq:crossover}
    \gamma = \frac{\gamma_{\rm ion} }{2} \frac{|\Omega|^2}{|\Omega|^2 + \kappa^2 / 8} = \frac{|\Omega|^2}{2\gamma_{\rm s}} \frac{|\Omega|^2}{|\Omega|^2 + \kappa^2 / 8},
\end{equation}
where we used Eq.~\eqref{eq:ioniz} for the ionization rate $\gamma_{\rm ion}$ [note that the denominator in Eq.~\eqref{eq:crossover} is different from that in Eq.~\eqref{eq:11-vs-00}]. For low drive amplitudes, $|\Omega|\ll\kappa$, this expression reduces to
\begin{equation}\label{eq:subrabi}
    \gamma = \frac{4 |\Omega|^4}{\gamma_{\rm s} \kappa^2} \propto P^2.
\end{equation}
For $|\Omega|\gg \kappa$, from Eq.~\eqref{eq:crossover} we recover Eq.~\eqref{eq:prep_under}, in which $\gamma \propto P$. 


\subsection{Effects of detuning\label{sec:detuning}}
The described process of the odd-state preparation has a \textit{resonant} character, i.e., it is the most effective when the drive frequency $\omega = (E_1 + E_2 + {\rm Re}\,\Sigma_{c})/\hbar$.  {Detuning of $\omega$ from the transition frequency rapidly suppresses the preparation rate.
To describe the suppression,} we focus on the case $\gamma_\mathrm{ion}\ll \kappa, |\Omega|$. The generalization of Eq.~\eqref{eq:11-vs-00} to the case of the drive detuned from the transition by $\delta \omega$ reads
\begin{equation}\label{eq:prob_rel_detuning}
    p_{11} = \frac{|\Omega|^2}{|\Omega|^2 + {\delta \omega^2} + \kappa^2 / 4}p_{00}.
\end{equation}
Using this relation in conjunction with Eq.~\eqref{eq:prob-decay}, we find for the inverse  odd-state preparation time:
\begin{equation}\label{eq:crossover_detuning}
    \gamma = \frac{|\Omega|^2}{2\gamma_{\rm s}} \frac{|\Omega|^2}{|\Omega|^2 + {\delta \omega^2 / 2}+ \kappa^2 / 8}.
\end{equation}
Similarly to the previously considered resonant case [Sec.~\ref{sec:res-P2-to-P}], the preparation rate crosses over from $\gamma\propto P^2$ at small drive power to $\gamma\propto P$ at higher drive power.
However, the detuning shifts the position of the crossover to a higher drive amplitude, $|\Omega| \sim \sqrt{\kappa^2 + {4\delta \omega^2}}$ instead of $|\Omega| \sim \kappa$.

\subsection{Preparation of the even state}
\label{sec:res-ion}
In the above we described how driving can be used to deterministically prepare the weak link in an odd state. 
A similar process can be used to prepare the system in the even state. To explain this, let us assume that the weak link is initialized in the odd state $|1_1, 0_2\rangle$, and a drive of frequency $\omega = (E_2 - E_1) / \hbar$ is applied. 
Absorption of a first drive photon promotes the quasiparticle from the first Andreev level to the second one. If the second level is sufficiently close to the continuum edge, see Eq.~\eqref{eq:condition_3},
then an absorption of a second photon ionizes the quasiparticle, bringing the weak link to an even state $\ket{0_1, 0_2}$ (with one quasiparticle in the continuum). In contrast to the preparation of the odd state, the preparation of the even state does not rely on the presence of residual Coulomb interaction in the weak link. This is because the odd-parity transition at $\omega = (E_2 - E_1) / \hbar$ is generally not resonant with any of the transitions in the even parity sector. 

The preparation of the even state can be described with the help of a phenomenological theory similar to the theory in Secs.~\ref{sec:res_gen}--\ref{sec:detuning}.
In particular, the dependence of the inverse preparation time on the drive amplitude is given by Eqs.~\eqref{eq:half_of_ion}, \eqref{eq:prep_over}, \eqref{eq:crossover}, and 
\eqref{eq:crossover_detuning}, in full similarity to the preparation of the odd state. 
The only difference
is in the value of the phenomenological parameters $|\Omega|$, $\gamma_{\rm s}$, and $\kappa$. To highlight that these parameters are different for the even-state preparation, in what follows we label them by
$|\widetilde{\Omega}|$, $\tilde{\gamma}_s$, and $\tilde{\kappa}$. As a reminder, coupling strength $|\widetilde{\Omega}|$ gives the Rabi frequency for the transition $\ket{1_1,0_2}\rightarrow\ket{0_1,1_2}$. Parameter $\tilde{\gamma}_s$ determines the ionization rate of the second Andreev level, $\gamma_{\rm ion} = |\widetilde{\Omega}|^2 / \tilde{\gamma}_s$.  {Expression for $\tilde{\gamma}_s$ is similar to Eq.~\eqref{eq:gamma_s_matrix_el}, 
\begin{equation}
\tilde{\gamma}_s^{-1} = \hbar\,\nu(E_2 + \hbar\omega) |\widetilde{\alpha}_c|^2.
\end{equation}
The only difference comes from the difference in frequency $\omega$. Here, $\omega = (E_2 - E_1) / \hbar$ while $\omega = (E_1 + E_2) / \hbar$ in Eq.~\eqref{eq:gamma_s_matrix_el}. $\widetilde{\alpha}_c$ is the dimensionless coupling at the former frequency.}
$\tilde{\kappa}$ gives the rate of a quasiparticle relaxation from the second level into the first level, $\ket{0_1, 1_2} \rightarrow \ket{1_1, 0_2}$ (as a reminder, for the odd state preparation $\kappa$ denoted the rate of quasiparticle recombination, $\ket{1_1, 1_2}\rightarrow \ket{0_1,0_2}$).

\subsection{Role of quasiparticle poisoning}
\label{sec:qps}
So far, we focused on the idealized case in which the parity state of the weak link is stable in the absence of driving. In this case, driving allows for the deterministic preparation of a desired parity state, as we explained.  {However, in recent experiments it was observed that the parity stochastically switches between even and odd; this was attributed to poisoning by the non-equilibrium quasiparticles~\cite{DiCarlo2013,Siddiqi2014,Serniak2019,Devoret2020}}. In the presence of stochastic parity switching, ideal deterministic preparation of the parity state is impossible. 
The fidelity of the state preparation is determined by how the preparation rate $\gamma$ compares to the parity switching rate, as we demonstrate below.

Let us denote the parity switching rates in the absence of the drive as $\Gamma_{\rm eo}$ and $\Gamma_{\rm oe}$ for the even-to-odd and odd-to-even transitions, respectively. For concreteness, we first focus on the preparation of the odd state. In this case, application of the resonant drive [in a way described in Sec.~\ref{sec:res_gen}] changes the rate of the even-to-odd switching from $\Gamma_{\rm eo}$ to $\Gamma_{\rm eo} + \gamma$ [see Eq.~\eqref{eq:crossover} for $\gamma$]. At the same time, the rate of odd-to-even switching remains unchanged, as long as the drive power is not too large. We can find the steady state probability of finding the system in the odd parity by applying a detailed balance condition
\begin{equation}
\label{eq:po}
    p_\mathrm{o} = \frac{\Gamma_{\rm eo} + \gamma}{\Gamma_{\rm eo} + \gamma + \Gamma_{\rm oe}}; 
\end{equation}
the even state probability is $p_{\rm e} = 1 - p_{\rm o}$.
If $\gamma\gg \Gamma_{\rm oe}, \Gamma_{\rm eo}$, then $p_\mathrm{o}$ becomes close to unity, $p_\mathrm{o} \approx 1 - \Gamma_{\rm oe}/\gamma$. A relation similar to Eq.~\eqref{eq:po} with $\mathrm{o} \leftrightarrow \mathrm{e}$ and $\gamma \leftrightarrow \tilde{\gamma}$ holds for the even state preparation.

It appears from Eq.~\eqref{eq:po} that increasing the drive power should bring the odd state probability $p_{\rm o}$ closer and closer to unity (since $\gamma$ increases with the drive power). This trend, however, breaks down in the high-power regime. As it was shown in Section~\ref{sec:res_gen}, in this regime ($|\Omega| \gg \gamma_{\rm s}$) the even-to-odd rate saturates at a value $\gamma \sim \gamma_{\rm s}$. This puts an upper limit on the fidelity of the odd state preparation,
\begin{equation}
    \label{eq:fidelity_sat}
    1 - p_{\rm o}^\mathrm{max} \sim \frac{\Gamma_{\rm oe}}{\gamma_{\rm s}},
\end{equation}
where $\Gamma_{\rm oe}$ is the rate of transitions from odd parity to even parity due to quasiparticle poisoning.
A similar limit holds for the preparation of the even state. We note that the limit imposed by Eq.~\eqref{eq:fidelity_sat} is loose and is unlikely to be reached experimentally (see comment at the end of Section~\ref{sec:res_gen}).


{
In the case of the odd-state preparation, Coulomb interaction puts an additional limit on the achievable fidelity.
Recall that to change the parity from even to odd, one drives the transition $\ket{0_1, 0_2}\rightarrow \ket{1_1, 1_2}$. The same drive evaporates a quasiparticle from the upper level bringing the system into the odd-parity ground state, $\ket{1_1, 0_2}$. 
For a sufficiently weak drive, the production of quasiparticles by the drive stops at this stage, thanks to the Coulomb blockade [see Fig.~\ref{fig:summary}]. 
We note, however, that a high-amplitude drive can break through the Coulomb blockade resulting in a transition $\ket{1_1, 0_2}\rightarrow \ket{2_1, 1_2}$ (even though it is detuned from resonance by an amount $\delta \omega = U / \hbar$ determined by the interaction strength). 
A subsequent evaporation of a quasiparticle from $\ket{2_1, 1_2}$ and recombination of the remaining two quasiparticles reverts the state back to $\ket{0_1, 0_2}$. Thus, the breakdown of the Coulomb blockade by the strong drive compromises the odd-state preparation fidelity.}

Quantitatively, the described effect can be taken into the account in Eq.~\eqref{eq:po} by replacing $\Gamma_{\rm oe}$ with $\Gamma_{\rm oe} + |\Omega|^2 \gamma_{\rm ion} / [2(U / \hbar)^2]$. Here, for $|\Omega|\lesssim U / \hbar$, factor $|\Omega|^2 / [2(U / \hbar)^2]$ determines the probability to find the system in state $\ket{2_1, 1_2}$ (assuming it starts in the odd state) and $\gamma_{\rm ion}$ is the rate of quasiparticle evaporation from state $\ket{2_1, 1_2}$ \footnote{We also assume that the rate of recombination $\kappa$ satisfies $U / \hbar \gg \kappa\gg \Gamma_{\rm oe}, \Gamma_{\rm eo}$. The second inequality guarantees that recombination is not the bottleneck process in the odd-to-even transition. The first inequality ensures the applicability of the replacement of $\Gamma_{\rm oe}$ by $\Gamma_{\rm oe} + |\Omega|^2 \gamma_{\rm ion}/[2 (U / \hbar)^2]$ used in deriving Eqs.~\eqref{eq:po-vs-U} and \eqref{eq:opt-omega}.}. With this replacement, it is apparent that when $|\Omega| \sim U / \hbar$ the fidelity of the odd state preparation becomes poor, $1 - p_{\rm o}\sim 1/2$. The fidelity is also poor when $|\Omega|= 0$. Therefore, the maximal fidelity is achieved at an intermediate drive strength. Explicitly we find
\begin{equation}
\label{eq:po-vs-U}
1 - p_\mathrm{o}^\mathrm{max} = \frac{(8\Gamma_{\rm oe}\gamma_{\rm s})^{1/2}}{U/\hbar},
\end{equation}
which is achieved at drive amplitude
\begin{equation}
\label{eq:opt-omega}
    |\Omega_{\rm opt}| = \left(2\Gamma_{\rm oe}\gamma_{\rm s}\right)^{1/4} \bigl( U / \hbar \bigr)^{1/2}.
\end{equation}
In deriving Eqs.~\eqref{eq:po-vs-U} and \eqref{eq:opt-omega} we assumed $\Gamma_{\rm oe} = \Gamma_{\rm eo}$ for simplicity, and we also assumed $(\Gamma_{\rm oe}\gamma_{\rm s})^{1/2}\ll U/\hbar$. 
 {According to Eq.~\eqref{eq:po-vs-U}, the maximum achievable fidelity increases with the increase of the interaction strength $U$, until the quasiparticle poisoning becomes the main constraint and the maximum fidelity is given by~\eqref{eq:fidelity_sat}. This} emphasizes that the interaction is instrumental for preparing the odd ground state with a high fidelity (in contrast to the preparation of the even ground state).

Which of the two restrictions [cf.~Eqs.~\eqref{eq:fidelity_sat} and \eqref{eq:po-vs-U}] limits the fidelity of the odd state preparation in practice depends on the comparison between $\Gamma_{\rm oe}$, $\gamma_{\rm s}$, and $U/\hbar$. This comparison is sensitive to microscopic details of the system.

\subsection{Discussion of the experiment}
The main predictions of our theory are consistent with a recent experimental work~\cite{jaap2021}.

A striking observation of Ref.~\onlinecite{jaap2021} is that a high-power drive, resonant with a transition in a given parity sector, brings the system into an \textit{opposite} parity sector. In particular,
driving the weak link at the frequency of the even parity transition brought it to the odd state. This is in agreement with our parity preparation mechanism [see Sec.~\ref{sec:res_gen}]. 
According to our theory, residual Coulomb interaction is required for the preparation of the odd state.
Signatures of Coulomb interaction in devices similar to that of Ref.~\onlinecite{jaap2021} were indeed recently reported~\cite{Fatemi2022, Urbina2022}.

The work \cite{jaap2021} also reports the dependence of parity preparation rates on the power of the applied drive. An attempt to fit the dependence with $P^\alpha$ resulted in exponent $\alpha$ between $1$ and $2$. Our theory predicts a crossover between $\gamma\propto P^2$ and $\gamma\propto P$ with the increase of power. This implies that the experimental data likely belongs to the crossover regime. According to our theory, the crossover happens when the Rabi rate compares to the linewidth, $|\Omega|\sim \kappa$ (unfortunately, $\kappa$ and $|\Omega|$ were not independently measured in Ref.~\onlinecite{jaap2021}).

Due to the quasiparticle poisoning, the parity preparation achieved in the experiment \cite{jaap2021} was not ideal.  {However, it was observed that the fidelity increases with the drive power $P$ reaching close-to-unity values at high $P$ (specifically, $0.89$ for the odd state preparation and $0.94$ for the even state preparation).} This observation is consistent with our theory, cf.~Eq.~\eqref{eq:po}.

\section{Microscopic model}\label{sec:micro}

In this section, we present a minimal microscopic model allowing one to evaluate the rates of preparation of the odd and even states, and find the dependence of these rates on relevant parameters such as the phase bias $\varphi$ across the junction.

We consider a weak link between the two superconducting leads which hosts two transport channels. 
The simplest and commonly used model describing such a configuration is that of a short junction. In this model, the two channels are independent of one another, and each gives rise to a single, spin-degenerate Andreev level.
For our purposes, however, the short junction model is insufficient.
Indeed, our mechanism relies on a process  {in which a drive} either breaks a Cooper pair into two quasiparticles  {belonging to} different levels (odd state preparation), or transfers a quasiparticle between the levels (even state preparation). Since the channels are independent in the short junction model, neither of these processes is allowed.
The independence of channels stems from neglecting the dwell time $\tau_{\rm dw}$ of a quasiparticle in the junction.
 {To quantify the state preparation rates, we take a short dwell time $\tau_{\rm dw} \ll \hbar/\Delta$ into the account.}

 \begin{figure}[t]
  \begin{center}
    \includegraphics[scale=1]{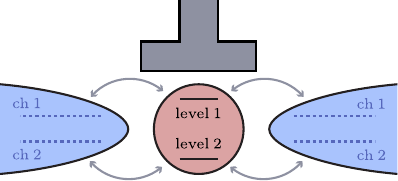}
    \caption{Schematic of a microscopic model. Quantum dot hosts two levels while each lead hosts two transport channels. The levels in the dot are tunnel coupled to the channels in the leads. In the main text, we consider a simplified situation in which the first level in the dot is only coupled to the first channel in each lead and, correspondingly, the second level is coupled to the second channel. The generic situation is considered in Appendix~\ref{sec:non-diagonalt}.}
    \label{fig:microm}
  \end{center}
\end{figure}

A model that allows for a systematic account of the finite dwell time is that of a quantum dot tunnel-coupled to two superconducting leads. The many-body Hamiltonian of the system is given by
\begin{equation}
    \label{eq:ham-micro}
    H = H_{\mathrm{qd}} + H_{\mathrm{sc}} + H_{\mathrm{tun}} + H_\mathrm{drive}(t).
\end{equation}
Here $H_{\mathrm{qd}}$ describes the levels on the dot:
\begin{equation}\label{eq:qd}
    H_{\rm qd} = c^\dagger \hat{\epsilon}_0 \tau_z c + H_{\rm C},
\end{equation}
where $c = (c_{1,\uparrow}, c_{2,\uparrow}, c_{1, \downarrow}^\dagger, c_{2, \downarrow}^\dagger)^T$ and $c_{\alpha, \sigma}$ is the annihilation operator of an electron in the level $\alpha = 1$ or $2$ on the dot, with spin $\sigma =\,\uparrow$ or $\downarrow$. Matrix $\hat{\epsilon}_0 = {\rm diag}\{\epsilon_1, \epsilon_2\}$ contains the energies of the two levels (computed with respect to the Fermi level). We assume the levels to be spin-degenerate. We denote the Pauli matrices in the Nambu space by $\tau_{j}$ with $j\in\{x, y, z\}$.  {In the considered limit of short dwell time, the specific values $\epsilon_1$ and $\epsilon_2$ will not be consequential for our results (assuming $\epsilon_\alpha \lesssim \Delta$).} $H_{\rm C}$ in Eq.~\eqref{eq:qd} describes the Coulomb interaction; we leave this term unspecified for the moment. We assume that the Coulomb interaction is well-screened so that it plays only a residual role: it  provides a mismatch between transition frequencies in the even and odd parity sectors [see Fig.~\ref{fig:summary} and Sec.~\ref{sec:koo}].

Term $H_{\rm sc}$ in Eq.~\eqref{eq:ham-micro} is the Hamiltonian of the superconducting leads. Focusing on two transport channels in each lead, we consider $H_{\rm sc}$ given by the following expression:
\begin{equation}\label{eq:H_sc}
    H_{\rm sc} = \sum_{i = R, L} \sum_\xi \psi^\dagger_{i, \xi} (\xi \tau_z + \Delta \tau_x) \psi_{i, \xi},
\end{equation}
where $\psi_{i,\xi} = (\psi_{i, \xi, 1,  \uparrow}, \psi_{i,\xi, 2, \uparrow}, \psi^\dagger_{i,\xi, 1, \downarrow}, \psi^\dagger_{i,\xi, 2, \downarrow})^T$. Operator $\psi_{i, \xi, \beta, \sigma}$ annihilates an electron with spin $\sigma$ in the lead $i =\,R$ or $L$. Index $\beta \in \{ 1, 2\}$ differentiates the two channels, whereas $\xi$ labels states in a given channel by their respective normal-state energies; $\xi = 0$ corresponds to the Fermi level.

The tunnel-coupling between the quantum dot and the leads is described in Eq.~\eqref{eq:ham-micro} by
\begin{equation}\label{eq:tun}
    H_{\rm tun} = \frac{1}{\sqrt{L}} \sum_{i = R,L} \sum_\xi c^\dagger \tau_z e^{i \varphi_i \tau_z / 2} \hat{t}_i \psi_{i,\xi} + {\rm h.c.},
\end{equation}
where $L$ is the normalization length for a channel, and $\varphi_i$ is the superconducting phase in the lead $i$. We fix the gauge in which $\varphi_L = \varphi$ and $\varphi_R = 0$. In Eq.~\eqref{eq:tun}, $\hat{t}_i$ is a $2 \times 2$ \textit{matrix} composed of tunneling amplitudes $t_{i, \alpha\beta}$ between channel $\beta$ in the lead $i$ and level $\alpha$ on the quantum dot. For simplicity, we assume $t_{i, \alpha\beta}$ to be diagonal (see Appendix~\ref{sec:non-diagonalt} for the discussion of the  {general} case). It is convenient to characterize the tunneling between the dot and the leads by the respective normal-state tunneling rates $\Gamma_{i, \alpha}  {/ \hbar} = \pi \nu_0 [\hat{t}_i \hat{t}_i^\dagger]_{\alpha\alpha}  {/ \hbar}$, where $\nu_0$ is the density of states in a given channel of a normal metal  {[we note that $\Gamma_{i,\alpha}$ have the units of energy]}.
The characteristic scale  $\Gamma  {/ \hbar}$  of the rates $\Gamma_{i,\alpha}  {/ \hbar}$ is related to the electron dwell time in the junction, $\tau_{\rm dw} = \hbar / \Gamma$.

Finally, the term $H_{\rm drive}(t)$ in Eq.~\eqref{eq:ham-micro} describes the microwave drive.
We assume throughout this section the driving is performed by applying an ac voltage to the gate, see Fig.~\ref{fig:summary}.
The respective term in the Hamiltonian is
\begin{equation}\label{eq:drive}
    H_\mathrm{drive}(t) = \cos (\omega t)\mathcal{E}_0 c^\dagger \hat{d} \tau_z c,
\end{equation}
where $\mathcal{E}_0$ is the amplitude of the electric field at the dot, and the dipole moment $\hat{d}$ is a $2 \times 2$ matrix acting in the subspace of levels on the dot.
In Section~\ref{sec:dissc}, we also discuss the case of the current drive, in which the ac voltage is applied between the superconducting leads.

\subsection{Static case}
To start with, we assume that the driving is absent and find the energy spectrum and the corresponding wavefunctions of the Hamiltonian \eqref{eq:ham-micro}.
With the Coulomb repulsion neglected, this amounts to solving the system of the Bogoliubov-de Gennes (BdG) equations [see Sec.~\ref{sec:koo} for a discussion of the Coulomb interaction effects]. The BdG equations read:
\begin{subequations}
\begin{align}
    (\xi \tau_z + \Delta \tau_x) \Psi_{i, \xi} + \frac{1}{\sqrt{L}} \tau_z e^{-i\frac{\varphi_i \tau_z}{2}}\hat{t}_i^\dagger C &= E \Psi_{i, \xi},\label{eq:bdg_1}\\
    \hat{\epsilon}_0 \tau_z C + \frac{1}{\sqrt{L}} \sum_{i = R, L}\sum_\xi \tau_z e^{i\frac{\varphi_i \tau_z}{2}}{\hat{t}_i}\Psi_{i,\xi} &= E C,\label{eq:bdg_2}
\end{align}
\end{subequations}
where $C$ and $\Psi_{i,\xi}$ are the components of the wave function in the dot and in the $i$-th lead, respectively.
To solve this system, we express $\Psi_{i,\xi}$ is terms of $C$ with the help of the first equation, and then substitute the result in the second equation. Then, upon performing the integration over $\xi$, we arrive to the following equation for $C$:
\begin{equation}\label{eq:bdg_for_C}
    E\left(1 + \frac{\sum_{i = R, L} \hat{\Gamma}_i}{\sqrt{\Delta^2 - E^2}}\right)C = \hat{\epsilon}_0\tau_z C + \hat{\gamma} C,
\end{equation}
where
\begin{equation}
    \hat{\gamma} = \frac{\Delta}{\sqrt{\Delta^2 - E^2}}\sum_{i = R, L}
    \begin{pmatrix}
    0 &  \hat{\Gamma}_i e^{i\varphi_i}\\
    \hat{\Gamma}_i e^{-i\varphi_i} & 0 
    \end{pmatrix}_{\hspace{-0.05cm}\tau}
\end{equation}
(subscript $\tau$ indicates that the matrix acts in the Nambu space).
In these equations, $\hat{\Gamma}_i  {/ \hbar} = \pi \nu_0 \hat{t}_i \hat{t}_i^\dagger  {/ \hbar}$ is the diagonal matrix of the tunneling rates introduced earlier. 

 \begin{figure*}[t]
  \begin{center}
    \includegraphics[scale = 1]{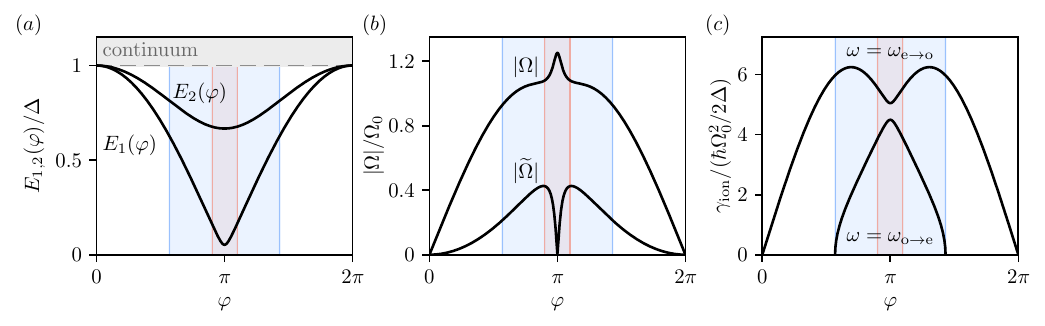}
    \caption{(a) Energies of the Andreev bound states as a function of phase bias $\varphi$ for a two-channel junction with $\Delta \cdot \tau_{\rm dw} / \hbar \ll 1$. (b) Coupling between the junction states due to the drive in even parity sector $\Omega$ and in the odd parity sector $\widetilde{\Omega}$. The coupling is normalized by $\Omega_0 \equiv ({\cal E}_0 |d_{12}| / \hbar) \cdot \Delta /(\Gamma_1 + \Gamma_2)$. (c) Ionization rate of the second Andreev level by a drive applied at $\omega_{\rm e\rightarrow o} = (E_1(\varphi) + E_2(\varphi)) / \hbar$ [preparation of the odd state], and at $\omega_{\rm o \rightarrow e} = (E_2(\varphi) - E_1(\varphi)) / \hbar$ [preparation of the even state]. The red stripe in all three panels depicts a domain of phases in which preparation of the odd state is possible, cf.~Eqs.~\eqref{eq:condition_1}, \eqref{eq:condition_2}; in the blue stripe, the even state can be prepared [cf.~Eq.~\eqref{eq:condition_3}]. Parameters are chosen as $\Gamma_{R,2} = 5\Gamma_{L, 2}$, $\Gamma_{L,1} = 1.5\Gamma_{L, 2}$, $\Gamma_{R,1} = 1.8\Gamma_{L,2}$,  $\Gamma_{L,2} = 5 \Delta$, and $|d_{22}|^2 / |d_{12}|^2  = 3$.}
    \label{fig:coupling}
    \label{fig:energies}
  \end{center}
\end{figure*}

Equation~\eqref{eq:bdg_for_C} allows one to find the spectrum of the Andreev levels and the respective wave functions 
for arbitrary $\Delta / \Gamma$. Since we are interested in the regime of a short dwell time, we focus on the case of $\Delta / \Gamma \ll 1$.
Under this condition, one can dispense with $1$ in the brackets on the left hand side of Eq.~\eqref{eq:bdg_for_C}, and with the first term on the right hand side [we assume that $|\hat{\epsilon}_0| \lesssim \Delta$].
Then, Eq.~\eqref{eq:bdg_for_C} simplifies to
\begin{equation}
\label{eq:zero-dwell}
    \Delta \sum_{i = R,L}\begin{pmatrix}
    0 & \hat{\Gamma}_i e^{i\varphi_i}\\
    \hat{\Gamma}_i e^{-i\varphi_i} & 0 \end{pmatrix}_{\hspace{-0.05cm}\tau} C = E\sum_{i = R,L} \hat{\Gamma}_i C.
\end{equation}
By assumption, matrices $\hat{\Gamma}_i$ are diagonal, which means that Eq.~\eqref{eq:zero-dwell} can be separately solved for each channel $\alpha \in \{1, 2\}$.
This yields
\begin{equation}\label{eq:energies}
    E_{\alpha}(\varphi) = \Delta \sqrt{1 - {\cal T}_\alpha \sin^2(\varphi / 2)},
\end{equation}
where 
\begin{equation}
    {\cal T}_\alpha = \frac{4\Gamma_{R,\alpha}\Gamma_{L,\alpha}}{\Gamma_\alpha^2},\quad \Gamma_\alpha =  \Gamma_{R,\alpha} + \Gamma_{L,\alpha}.
\end{equation}

The components of the wavefunctions on the dot are given by
\begin{align}
\label{eq:solution}
    C_{\alpha, +} =\,&\frac{{\cal N}_\alpha(\varphi)}{\sqrt{2}}
    \begin{pmatrix}
        1\\
        e^{-iz_\alpha(\varphi)}
    \end{pmatrix}_{\hspace{-0.05cm}\tau},\notag\\
    &\hspace{0.5cm}e^{-iz_\alpha(\varphi)} = \frac{\Delta}{E_\alpha(\varphi)}\frac{\Gamma_{L,\alpha} e^{-i\varphi} + \Gamma_{R,\alpha}}{\Gamma_{\alpha}}.
\end{align}
Here factor ${\cal N}_\alpha(\varphi)$ satisfies \cite{Kurilovich2021}
\begin{equation}
\label{eq:norm}
{\cal N}_\alpha^2(\varphi) = \frac{\sqrt{\Delta^2 - E^2_\alpha(\varphi)}}{\Gamma_{\alpha}}.
\end{equation}
This factor describes spreading of the wavefunction from the dot into the leads. The spreading increases with the increase of $\Gamma_\alpha$, or when the energy of the level approaches the continuum, $\varphi \rightarrow 0$.
We note that Eqs.~\eqref{eq:energies}--\eqref{eq:norm} break down in the vicinity of $\varphi = 0$ of width $\delta \varphi \sim \Delta / \Gamma$. In particular, $E_\alpha(0)$ are in fact separated from $E = \Delta$ by $\delta E \sim \Delta^3 / \Gamma^2$.

Equation~\eqref{eq:solution} gives a solution of the BdG equations with the positive energy (as indicated by a $+$ subscript). We will also need a respective negative energy solution with $E = -E_\alpha(\varphi)$. Its wavefunction is given by $C_{\alpha, -} = \tau_z C_{\alpha, +}$.

An example of the phase-dependence of the Andreev level energies is shown in Fig.~\ref{fig:energies}. The red stripe depicts an interval of phases in which the deterministic preparation of the even state is possible, cf.~condition \eqref{eq:condition_3}.
A more narrow blue stripe shows an interval of phases in which the odd-state preparation is allowed by conditions \eqref{eq:condition_1} and \eqref{eq:condition_2}.

\subsection{Dynamics}\label{sec:dynamics}
Next, we assume that a drive is applied to the gate adjacent to the quantum dot, see Eq.~\eqref{eq:drive}. We consider two cases: either the drive frequency is resonant with a transition in the even sector, $\hbar\omega = E_1(\varphi) + E_2(\varphi)$, or with a transition in the odd sector, $\hbar \omega = E_2(\varphi) - E_1(\varphi)$. For both of these cases, we evaluate the coupling $\Omega \equiv \Omega(\varphi)$ entering the phenomenological theory, cf.~Eq.~\eqref{eq:Vt}. We also compute the rate of ionization of the Andreev levels $\gamma_\mathrm{ion} \equiv \gamma_{\rm ion}(\varphi)$, cf.~Eq.~\eqref{eq:ioniz}.
 
\subsubsection{Evaluation of the coupling $\Omega(\varphi)$}
We start by computing the coupling $\Omega$ for the even transition at $\hbar\omega = E_1 + E_2$. It can be expressed as a matrix element of the drive operator:
\begin{equation}
    \hbar\Omega = \frac{\mathcal{E}_0}{\sqrt{2}} C_{2,+}^\dagger  \hat{d} \tau_z C_{1,-},
\end{equation}
where $C_{\alpha,\pm}$ are given by Eq.~\eqref{eq:solution}. Substituting these expressions, we find for the magnitude of the coupling:
\begin{equation}
\label{eq:rabi-1}
    \frac{\hbar|\Omega|}{\mathcal{E}_0 |d_{12}|} =  \frac{[(\Delta^2 - E_1^2)(\Delta^2-E_2^2)]^{\frac{1}{4}}}{\sqrt{2\,\Gamma_1\Gamma_2}} 
    \Bigl| \cos\Bigl(\frac{z_1 - z_2}{2}\Bigr)\Bigr|,
\end{equation}
where we suppressed the phase arguments of $E_\alpha$ and $z_\alpha$ [$z_\alpha$ is defined in Eq.~\eqref{eq:solution}].
Coupling $\widetilde{\Omega}$  for the odd state transition at $\hbar \omega = E_2 - E_1$ can be obtained in the same way. We find
\begin{equation}
\label{eq:rabi-2}
    \frac{\hbar|\widetilde{\Omega}|}{\mathcal{E}_0 |d_{12}|} =  \frac{[(\Delta^2 - E_1^2)(\Delta^2-E_2^2)]^{\frac{1}{4}}}{2\sqrt{\Gamma_1\Gamma_2}} 
    \Bigl| \sin\Bigl(\frac{z_1 - z_2}{2}\Bigr)\Bigr|.
\end{equation}

In both cases the coupling can be estimated as
\begin{equation}\label{eq:coupling}
 |\Omega|,|\widetilde{\Omega}| \sim \frac{1}{\hbar}\mathcal{E}_0 |d_{12}| \frac{\Delta}{\Gamma}.
\end{equation}
Here we assumed $\hbar\omega, E_1, E_2 \sim \Delta$ and denoted  {the characteristic value of the tunneling rate by $\Gamma / \hbar$.}
The coupling is attenuated in comparison with  {the ``bare'' dipole coupling} ${\cal E}_0 |d_{12}|$ by a factor $\sim \Delta/\Gamma \ll 1$. This factor originates from spreading of the wavefunction from the dot into the leads, where the electric field produced by the gate is screened. 

The phase dependence of couplings $\Omega$ and $\widetilde{\Omega}$ is depicted in Fig.~\ref{fig:energies}. Its character near $\varphi = \pi$ is of a particular note. Specifically, one of the two couplings vanishes at $\varphi = \pi$ due to destructive particle-hole interference. Which of the two processes this happens for depends on the relation between the tunneling rates. If $(\Gamma_{1,R} - \Gamma_{1,L})\cdot (\Gamma_{2,R} - \Gamma_{2,L})> 0$, then $z_1(\pi) = z_2(\pi) = 0$ [cf.~Eq.~\eqref{eq:solution}] and thus $|\widetilde{\Omega}| = 0$, as shown in Fig.~\ref{fig:coupling}. In the opposite case, $z_1(\pi) - z_2(\pi) = \pm\pi$ and $|\Omega| = 0$. This cancellation occurs only in the leading order in the dwell time.

We note that Eqs.~\eqref{eq:rabi-1} and \eqref{eq:rabi-2} break down in the vicinity of zero phase of width $\delta \varphi \sim \Delta / \Gamma \ll 1$, where energies \eqref{eq:energies} poorly approximate the exact solution of Eq.~\eqref{eq:bdg_for_C}. There, Eqs.~\eqref{eq:rabi-1} and \eqref{eq:rabi-2} underestimate the coupling strength.

\subsubsection{Evaluation of the ionization rate $\gamma_{\rm ion}(\varphi)$}
Next, we compute the rate of ionization of the upper Andreev level by the applied drive, cf.~Eq.~\eqref{eq:ioniz}. Using Fermi's Golden rule, we obtain the following expression for the rate:
\begin{equation}
    \gamma_{\rm ion} = \frac{2\pi}{\hbar} \sum_c |\langle 1_{c} | \frac{1}{2}\mathcal{E}_0 \hat{d} \tau_z |1_2 \rangle|^2 \delta(E_2 + \hbar\omega - E_c).
\end{equation}
Here $\ket{1_2}$ denotes a state with a quasiparticle in the upper Andreev level; $\ket{1_{c}}$ is a state with a quasiparticle of energy $E_c$ in the continuum.
The sum over the final states can be carried out by introducing the Green's function of the system, $G(E)$. In fact, the only component of the Green's function that is relevant for the evaluation of $\gamma_{\rm ion}$ is $G_{\rm dd}$, i.e., the component describing the dot. We find:
\begin{equation}\label{eq:ion_through_GF}
    \gamma_{\rm ion} = -\frac{2}{\hbar}\frac{\mathcal{E}^2_0}{4}\mathrm{Im}
     \langle 1_2| \hat{d} \tau_z G_{\rm dd}(\hbar \omega + E_2) \hat{d} \tau_z |1_2 \rangle.
\end{equation}
In the limit of short dwell time, $\Delta / \Gamma \ll 1$,
$G_{\rm dd}$ can be expressed as \cite{Kurilovich2021}
\begin{equation}\label{eq:dot_GF}
    {G}_{\rm dd}(E) =  \frac{-i\sqrt{E^2 - \Delta^2}}{\hat{\Gamma}_R + \hat{\Gamma}_L}\frac{1}{E - H_0},
\end{equation}
where the effective Hamiltonian of the Andreev levels is given by
\begin{equation}\label{eq:eff_H}
    H_0 = \frac{\Delta}{\hat{\Gamma}_R + \hat{\Gamma}_L} \sum_{i = R, L}\begin{pmatrix}
    0 & \hat{\Gamma}_i e^{i\varphi_i}\\
    \hat{\Gamma}_i e^{-i\varphi_i} & 0 \end{pmatrix}_{\hspace{-0.05cm}\tau}.
\end{equation}
Using Eqs.~\eqref{eq:dot_GF} and \eqref{eq:eff_H} in Eq.~\eqref{eq:ion_through_GF}, we find
\begin{widetext}
\begin{equation}
\label{eq:ionizmicro}
    \gamma_\mathrm{ion} = \frac{\mathcal{E}_0^2}{2\hbar}\sqrt{\left(\hbar\omega+E_2\right)^2 - \Delta^2}\sqrt{\Delta^2 - E_2^2}\left(\frac{|d_{12}|^2 }{\Gamma_1 \Gamma_2}\frac{\cos^2\bigl(\frac{z_1 - z_2}{2}\bigr)}{\hbar\omega + E_2 - E_1}+\frac{|d_{12}|^2 }{\Gamma_1 \Gamma_2}\frac{\sin^2\bigl(\frac{z_1 - z_2}{2}\bigr)}{\hbar\omega + E_2 + E_1}+\frac{|d_{22}|^2}{\Gamma_2^2}\frac{1}{\hbar\omega + 2 E_2}\right).
\end{equation}
\end{widetext}
The square root factors here reflect the behavior of the local density of states at the weak link \cite{Glazman2013, Kurilovich2021}.  {We remind one that $\hbar \omega = E_1 + E_2$ in the case of the odd state preparation, and $\hbar \omega = E_2 - E_1$ in the case of the preparation of the even state. An example of the dependence of $\gamma_{\rm ion}$ on phase $\varphi$ is shown in Fig.~\ref{fig:energies}(c).}

Using Eq.~\eqref{eq:ionizmicro} we can estimate the rate $\gamma_{\rm ion}$ as
\begin{equation}\label{eq:ionrate}
    \gamma_{\rm ion} \sim \frac{1}{\hbar} \mathcal{E}_0^2 |d_{12}|^2 \frac{\Delta}{\Gamma^2},
\end{equation}
where we made assumptions that $\hbar\omega, E_1, E_2 \sim \Delta$ and $|d_{22}|\sim |d_{12}|$ and denoted the typical tunneling rate by  {$\Gamma / \hbar$}. 

We can combine Eqs.~\eqref{eq:coupling} and \eqref{eq:ionrate} to estimate the saturation rate $\gamma_{\rm s} = |\Omega|^2 / \gamma_{\rm ion}$ as
\begin{equation}\label{eq:saturation}
    \gamma_{\rm s} \sim \Delta / \hbar
\end{equation}
[the estimate is the same for $\tilde{\gamma}_s$].

\subsection{Coulomb interaction}
\label{sec:koo}
As explained in Section~\ref{sec:intro}, deterministic preparation of a state with a single quasiparticle is allowed by a residual Coulomb interaction in the weak link.
The interaction ensures that, the drive ceases to add more quasiparticles to the Andreev levels once the odd state is reached, see Fig.~\ref{fig:summary}. The blocking happens because interaction detunes the transition frequency of $\ket{1_1,0_2} \rightarrow \ket{2_1,1_2}$  {away from} that of $\ket{0_1,0_2} \rightarrow \ket{1_1,1_2}$ (two frequencies coincide in the absence of interaction). Here we find the magnitude of the detuning between the transitions, which we label as $U/\hbar$.

We begin by specifying a concrete form of the interaction Hamiltonian $H_C$ in Eq.~\eqref{eq:qd}. For simplicity, we assume that the interaction is determined by the total number $N$ of electrons on the dot only \footnote{Subtraction of $2$ from the number operator $N$ is introduced for convenience; it endows the Hamiltonian with the particle-hole symmetry. In fact, Eq.~\eqref{eq:coulomb_ham} is equivalent to $U_C N^2$ up to an inconsequential renormalization of the single-particle energies.}:
\begin{equation}\label{eq:coulomb_ham}
    H_C = U_C (N - 2)^2 = U_C (c^\dagger \tau_z c)^2,
\end{equation}
where the charging energy $U_C = e^2 / C$ is determined by the capacitance $C$ of the dot.

To find $U$, we perturbatively compute the corrections to the energies of the relevant states (i.e., $\ket{0_1,0_2}$, $\ket{1_1,1_2}$, $\ket{1_1,0_2}$, $\ket{2_1,1_2}$) due to the Coulomb interaction assuming $\Gamma \gg \Delta$. The general expressions for these corrections are bulky; we present them in  Appendix~\ref{app:coulomb}. The detuning between the considered transitions, however, admits a simple representation. We find
\begin{equation}
    \label{eq:delta-f}
    U = - 2 U_C \frac{\Delta^2 - E_1^2}{\Gamma^2_1}.
\end{equation}
Notably, the detuning is small compared to the ``bare'' value of the Coulomb repulsion. 
The weakening of the repulsion is the result of the extension of the Andreev states into the leads (where the interaction is well-screened).
 {Another notable feature of Eq.~\eqref{eq:delta-f} is that---in the considered model---$U$ is determined only by the properties of the channel hosting the lower Andreev level.}

\section{Discussion}
\label{sec:dissc}

{\bf Driving by phase difference.} Above, we considered a microwave drive applied to the gate adjacent to the weak link. Another commonly used approach is driving by an ac phase bias across the link. Our phenomenological theory also applies in this case. Here, we estimate the parameters $|\Omega|$, $\gamma_{\rm ion}$, and $\gamma_{\rm s}$  {entering it} within the microscopic model of Sec.~\ref{sec:micro}.

Let us start with the coupling strength $\Omega$ (we recall that this parameter directly gives the frequency of Rabi oscillations, $\Omega_{\rm R} = \Omega$). In direct analogy to a respective estimate for the gate drive, see Eq.~\eqref{eq:coupling}, we find
\begin{equation}\label{eq:couplingphase}
    \Omega \sim \frac{1}{\hbar} \Delta\,\delta \varphi \frac{\Delta}{\Gamma},
\end{equation}
where $\delta \varphi$ is the drive amplitude.

Same as in Eq.~\eqref{eq:coupling}, the coupling is suppressed by a small parameter $\Delta\,\tau_{\rm dw} / \hbar$. Notably, the ionization rate by the phase drive does not contain this small factor~\cite{Glazman2013}:
\begin{equation}\label{eq:ionphase}
    \gamma_{\rm ion} \sim \frac{1}{\hbar} \Delta (\delta \varphi)^2.
\end{equation}
This is strikingly different from Eq.~\eqref{eq:ionrate}.

We can use Eqs.~\eqref{eq:couplingphase} and~\eqref{eq:ionphase} to estimate the saturation rate $\gamma_{\rm s} = \Omega^2 / \gamma_{\rm ion}$. We obtain
\begin{equation}\label{eq:satphase}
    \gamma_{\rm s} \sim \frac{\Delta}{\hbar} \Bigl(\frac{\Delta}{\Gamma}\Bigr)^2.
\end{equation}
Clearly, the saturation rate is parametrically smaller for a phase drive than it is for the gate drive, cf.~Eqs.~\eqref{eq:saturation} and~\eqref{eq:satphase}. Therefore, it should be easier to observe the saturation of the odd-state preparation rate in the former case.

\begin{figure}[t]
  \begin{center}
    \includegraphics[scale = 1]{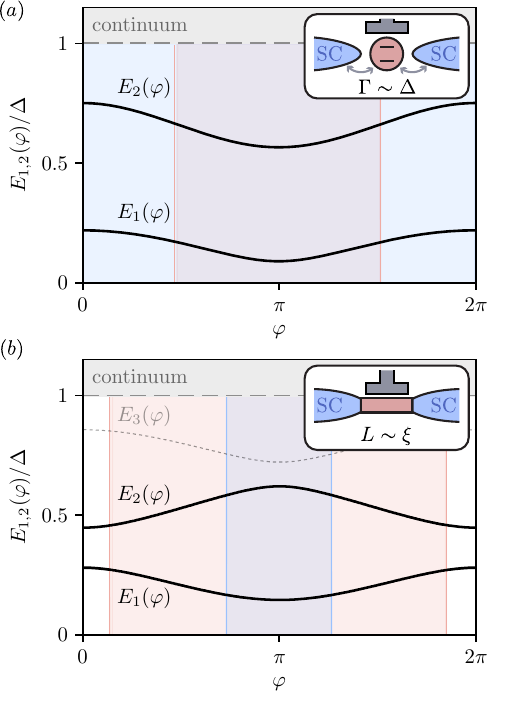}
    \caption{
    For a final dwell time $\tau_\mathrm{dw} \sim \hbar / \Delta$,  {preparation of the odd-parity state} can be achieved in a broad range of phase biases $\varphi$. This is in contrast to the regime $\tau_\mathrm{dw} \ll \hbar / \Delta$ considered in the main text, where  {the odd-state preparation} is only allowed close to $\varphi = \pi$  {[see Fig.~\ref{fig:energies}(a)]}. (a)  {Example of an energy} spectrum of a two-channel quantum dot coupled to superconducting leads in the regime $\tau_\mathrm{dw} \sim \hbar / \Delta$ (which is achieved at $\Gamma \sim \Delta$). Red (blue) region shows  {an interval of phases in which preparation of the odd (even) state} is allowed. (b) Regime $\tau_\mathrm{dw} \sim \hbar / \Delta$ can be also achieved in a single-channel weak link if the length $L$ of the link is comparable to the superconducting coherence length $\xi$. The panel shows the spectrum of such a ``long'' junction calculated within the framework of Refs.~\onlinecite{Yeyati2017,Krogstrup2019} {[we use $L = 3.5\xi$, $x_{0} = 0.75 L$, $\tau = 0.6$, and $\lambda_1 = \lambda_2$ in the notations of Ref.~\onlinecite{Krogstrup2019}]}. The meaning of red and blue regions is simlar to that in panel (a).}
    \label{fig:finite_dwell_time}
  \end{center}
\end{figure}

{\bf Finite dwell time.} Throughout our work, we made an assumption that the dwell time in the weak link is short, $\tau_\mathrm{dw} \ll \hbar/\Delta$. Under this assumption,
the preparation of the odd state is only allowed in a narrow interval of phases around $\varphi = \pi$, see Fig.~\ref{fig:coupling}. 
The narrowness of the interval results from a strong dispersion of the energy levels with phase.
 {Long dwell time, $\tau_\mathrm{dw} \gtrsim \hbar/\Delta$, suppresses the dispersion; the odd state preparation may now be allowed in a broad phase interval $\Delta\varphi\sim 1$.} We demonstrate this by numerically finding the energy spectrum in a quantum dot model with two channels and coupling $\Gamma \sim \Delta$ [see Fig.~\ref{fig:finite_dwell_time}(a)]. Additionally, we demonstrate that odd state can be prepared in a broad range of phases in a single-channel nanowire Josephson junction of length $L \sim \xi$ [see Fig.~\ref{fig:finite_dwell_time}(b)].




{\bf Spin-orbit interaction.}
In the absence of spin-orbit interaction, Coulomb interaction was necessary to deterministically  {prepare the odd-parity state}. The role of interaction was to remove degeneracy in the transition frequencies between the even and odd parity sectors, see Section~\ref{sec:koo}. Away from $\varphi = 0,\pi$, spin-orbit interaction is also sufficient to remove the degeneracy, even in the absence of Coulomb interaction. In this case, the fidelity of the odd state preparation is limited by the dynamics of spin-relaxation, careful treatment of which is beyond the scope of our manuscript.

\section{Conclusions}
\label{sec:concl}
We proposed a mechanism to deterministically prepare a weak link hosting two Andreev levels in  {the odd-parity state}, i.e., a state with a single quasiparticle trapped in the lowest Andreev level. The mechanism relies on driving the even transition with frequency $\hbar\omega = E_1 + E_2$, and on the residual Coulomb interaction between the quasiparticles. First, the drive breaks a Cooper pair generating two quasiparticles, one in each Andreev level. Then, if the transition frequency is high enough [cf.~Eq.~\eqref{eq:condition_1}], the same drive evaporates the quasiparticle from the upper level into the continuum thus leaving the weak link in the desired odd state. The presence of residual Coulomb interaction in the weak link prevents the drive from adding more quasiparticles once the odd state is reached, see Fig.~\ref{fig:summary}.

For this mechanism, we calculated the dependence of the rate $\gamma$ at which the odd state is prepared on the power $P$ of the applied drive. We showed that at small powers the rate scales as $\gamma\propto P^2$, see Eq.~\eqref{eq:subrabi}, crossing over to $\gamma \propto P$ when the drive-mediated coupling strength becomes comparable to the linewidth of the resonance at $\hbar\omega = E_1 + E_2$, see Eq.~\eqref{eq:prep_under}. At even higher powers, the rate saturates and ceases to increase with $P$, see Eq.~\eqref{eq:prep_over}. In the presence of quasiparticle poisoning, this puts a limit on the achievable fidelity of the odd state preparation, cf.~Eq.~\eqref{eq:fidelity_sat}.

A two-photon process similar to the one described above can be used to bring the weak link from the odd state to the even state. In contrast to the odd-state preparation, the preparation of the even state does not rely on residual Coulomb interaction.

We support our phenomenological theory with a minimal microscopic model of the weak link. The model consists of a quantum dot hosting two levels coupled to two superconducting leads and driven by applying a voltage to an adjacent gate. Within this model we evaluate the rates of odd- and even-state preparation and determine their dependence on the phase bias $\varphi$ across the weak link.

Our results explain a recent experiment \cite{jaap2021}. There, a microwave tone was applied to a weak link with a desire to drive the transitions in the charge-even parity sector. However, the link tended to switch to the opposite charge-parity sector upon the increase of the drive power. We attribute this behavior to our parity preparation mechanism. The power dependence of the parity switching rates in Ref.~\onlinecite{jaap2021} is qualitatively consistent with our results.
\acknowledgements{
We thank Jaap J. Wesdorp, Valla Fatemi, and Alfredo Levy Yeyati for insightful discussions. 
This research was sponsored by the Army Research Office (ARO) under grant number W911NF-22-1-0053, by the Office of Naval Research (ONR) under award number N00014-22-1-2764, by the NSF Grant No. DMR-2002275, and by the U.S. Department of Energy, Office of Science, National Quantum Information Science Research Centers, Co-design Center for Quantum Advantage (C2QA) under contract number DE-SC0012704. The views and conclusions contained in this document are those of the authors and should not be interpreted as representing the official policies, either expressed or implied, of the U.S. Government. The U.S. Government is authorized to reproduce and distribute reprints for Government purposes notwithstanding any copyright notation herein. A.S. and W.B. were supported by the Deutsche Forschungsgemeinschaft (DFG; German Research Foundation) via projects number 465140728, 467596333 and 425217212 (SFB 1432).}
\appendix
\begin{widetext}
\section{Corrections to the spectrum due to the Coulomb interaction}\label{app:coulomb}

Here, we calculate the first-order corrections to the energies of the relevant many-body states due to Coulomb interaction.  {It is these corrections that make the deterministic preparation of the odd-parity state possible. Indeed, the corrections create a mismatch in the transition frequencies between the even and odd parity states} and thus halt the quasiparticle addition once the odd state is reached [see Fig.~\ref{fig:summary}(b)]. We carry out the calculations within the microscopic model of Section~\ref{sec:micro}, i.e., we treat the weak link as a quantum dot hosting two Andreev levels. For simplicity, we focus on the regime of strong coupling to the leads, $\Gamma\gg\Delta$, and assume that the two-levels belong to two \textit{independent} transport channels. While both assumptions can be straightforwardly lifted, the corresponding calculation is beyond the scope of the present work.
 {We also} take the simplest possible interaction Hamiltonian of the form
\begin{equation}
\label{eq:hint}
    H_\mathrm{int} = U_C (N - 2)^2,\quad N = \sum_{\beta=1,2}\left(c^\dagger_{\beta,\uparrow} c_{\beta,\uparrow} + c_{\beta,\downarrow}^\dagger c_{\beta,\downarrow}\right).
\end{equation}
Here $-2$ term in $(N-2)^2$ is added to enforce particle-hole symmetry. Hamiltonian \eqref{eq:hint} neglects the exchange interaction between the quasiparticles as well as the fine structure due to the spin-orbit coupling~\cite{Urbina2022}.

We begin by expanding the dot operators $c_{\beta, \uparrow}$ and $c_{\beta, \downarrow}^\dagger$ through the eigenstate operators of the non-interacting problem \cite{Kurilovich2021}:
\begin{equation}
    \label{eq:decomps}
    c_{\beta,\uparrow} = \sum_{|\epsilon|<\Delta} p_{\beta\epsilon} \gamma_{\beta\epsilon} + \sum_{|\epsilon|>\Delta} p_{\beta\epsilon} \gamma_{\beta\epsilon},\quad c_{\beta,\downarrow}^\dagger = \sum_{|\epsilon|<\Delta} h_{\beta\epsilon} \gamma_{\beta\epsilon} + \sum_{|\epsilon|>\Delta} h_{\beta\epsilon} \gamma_{\beta\epsilon}.
\end{equation}
Next, we substitute these decompositions into the interaction Hamiltonian~\eqref{eq:hint}. In doing that, we neglect the terms which contain the operators of the supgap levels either one or three times. These terms can only lead to corrections of the second order in $U_C$ to the transition frequencies between the discrete states. We also disregard the terms which do not contain the operators of the subgap levels since they do not affect the discrete part of the spectrum. As a result, we arrive at
\begin{gather}
H_\mathrm{int} = U_C \left(
\sum_{\beta}\sum_{|\epsilon|<\Delta}\sum_{|\epsilon^{\prime}|<\Delta}(p_{\epsilon \beta}^{\star}p_{\epsilon^{\prime}\beta}-h_{\epsilon^{\prime}\beta}h_{\epsilon \beta}^{\star})\gamma_{\epsilon \beta}^{\dagger}\gamma_{\epsilon^{\prime}\beta}\sum_{\delta}\sum_{|E|<\Delta}\sum_{|E^{\prime}|<\Delta}(p_{E\delta}^{\star}p_{E^{\prime}\delta}-h_{E^{\prime}\delta}h_{E\delta}^{\star})\gamma_{E\delta}^{\dagger}\gamma_{E^{\prime}\delta}+\right.\notag\\
+\sum_{\beta}\sum_{|\epsilon|<\Delta}\sum_{|\epsilon^{\prime}|<\Delta}(p_{\epsilon \beta}^{\star}p_{\epsilon^{\prime}\beta}-h_{\epsilon^{\prime}\beta}h_{\epsilon \beta}^{\star})\gamma_{\epsilon \beta}^{\dagger}\gamma_{\epsilon^{\prime}\beta}\sum_{\delta}\sum_{E<-\Delta}(|p_{E\delta}|^{2}-|h_{E\delta}|^{2})+\notag\\
+\sum_{\beta}\sum_{\epsilon<-\Delta}(|p_{\epsilon \beta}|^{2}-|h_{\epsilon \beta}|^{2})\sum_{\delta}\sum_{|E|<\Delta}\sum_{|E^{\prime}|<\Delta}(p_{E\delta}^{\star}p_{E^{\prime}\delta}-h_{E^{\prime}\delta}h_{E\delta}^{\star})\gamma_{E\delta}^{\dagger}\gamma_{E^{\prime}\delta}\notag\\
+\sum_{\beta}\sum_{|\epsilon|<\Delta}\sum_{|\epsilon^{\prime}|>\Delta}(p_{\epsilon \beta}^{\star}p_{\epsilon^{\prime}\beta}-h_{\epsilon \beta}^{\star}h_{\epsilon^{\prime}\beta})\gamma_{\epsilon \beta}^{\dagger}\gamma_{\epsilon^{\prime}\beta}\sum_{\delta}\sum_{|E^{\prime}|>\Delta}\sum_{|E|<\Delta}(p_{E^{\prime}\delta}^{\star}p_{E\delta}-h_{E^{\prime}\delta}^{\star}h_{E\delta})\gamma_{E^{\prime}\delta}^{\dagger}\gamma_{E\delta}+\notag\\
\left.
+\sum_{\beta}\sum_{|\epsilon^{\prime}|>\Delta}\sum_{|\epsilon|<\Delta}(p_{\epsilon^{\prime}\beta}^{\star}p_{\epsilon \beta}-h_{\epsilon^{\prime}\beta}^{\star}h_{\epsilon \beta})\gamma_{\epsilon^{\prime}\beta}^{\dagger}\gamma_{\epsilon \beta}\sum_{\delta}\sum_{|E|<\Delta}\sum_{|E^{\prime}|>\Delta}(p_{E\delta}^{\star}p_{E^{\prime}\delta}-h_{E\delta}^{\star}h_{E^{\prime}\delta})\gamma_{E\delta}^{\dagger}\gamma_{E^{\prime}\delta}\right).\label{eq:long-hint}
\end{gather}
Here, we used the normalization condition as well as particle-hole symmetry \cite{Kurilovich2021}
\begin{equation}
    \label{eq:ph}
    \sum_{\epsilon<0}\left(|p_{\epsilon\beta}|^2 + |h_{\epsilon\beta}|^2\right) = 1, \quad \sum_{\epsilon > \Delta} |h_\epsilon|^2 = \sum_{\epsilon < -\Delta} |p_\epsilon|^2.
\end{equation}
Next, we use a relation
\begin{equation}
    \label{eq:orthog}
    p_{\epsilon\beta}^\star p_{\epsilon^\prime \beta} - h_{\epsilon\beta}^\star h_{\epsilon^\prime \beta} = \alpha_\beta \delta_{\epsilon, -\epsilon^\prime},\quad \alpha_\beta = \sqrt{\Delta^2 - E_\beta^2} / \Gamma_\beta.
\end{equation}
which follows directly from the wave-functions given by Eq.~\eqref{eq:solution} in the limit $\Gamma \gg \Delta$. Eq.~\eqref{eq:orthog} allows us to neglect the terms in the second and in the third line of Eq.~\eqref{eq:long-hint}. Then, combining the terms in the final two lines of Eq.~\eqref{eq:long-hint}, we obtain
\begin{gather}
H_\mathrm{int} = U_C \left( \alpha_{1}^{2}\gamma_{+1}^{\dagger}\gamma_{+1}\gamma_{-1}\gamma_{-1}^{\dagger}+\alpha_{1}^{2}\gamma_{-1}^{\dagger}\gamma_{-1}\gamma_{+1}\gamma_{+1}^{\dagger} + 
\alpha_{2}^{2}\gamma_{+2}^{\dagger}\gamma_{+2}\gamma_{-2}\gamma_{-2}^{\dagger}+\alpha_{2}^{2}\gamma_{-2}^{\dagger}\gamma_{-2}\gamma_{+2}\gamma_{+2}^{\dagger}+ \notag\right.\\\left.
+\sum_\beta \sum_{|\epsilon|<\Delta} \left[\sum_{\epsilon^{\prime}>\Delta}(p_{\epsilon^{\prime}\beta}^{\star}p_{\epsilon \beta}-h_{\epsilon^{\prime}\beta}^{\star}h_{\epsilon \beta})(p_{\epsilon \beta}^{\star}p_{\epsilon^{\prime}\beta}-h_{\epsilon \beta}^{\star}h_{\epsilon^{\prime}\beta})
- \sum_{\epsilon^{\prime}<-\Delta}(p_{\epsilon \beta}^{\star}p_{\epsilon^{\prime}\beta}-h_{\epsilon \beta}^{\star}h_{\epsilon^{\prime}\beta})(p_{\epsilon^{\prime}\beta}^{\star}p_{\epsilon \beta}-h_{\epsilon^{\prime}\beta}^{\star}h_{\epsilon \beta})\right]\gamma_{\epsilon \beta}^{\dagger}\gamma_{\epsilon \beta}\right).\label{eq:not-so-long-hint}
\end{gather}
Next, we use Eq.~\eqref{eq:ph} and \eqref{eq:orthog}, together with another relation following from the particle-hole symmetry \cite{Kurilovich2021},
\begin{equation}
\sum_{\epsilon>\Delta} p_{\epsilon\beta} h_{\epsilon\beta}^\star = -\sum_{\epsilon<-\Delta} p_{\epsilon\beta} h_{\epsilon\beta}^\star. 
\end{equation}
This allows us to rewrite Eq.~\eqref{eq:not-so-long-hint} as
\begin{gather}
H_\mathrm{int} = U_C \left( \alpha_{1}^{2}\gamma_{+1}^{\dagger}\gamma_{+1}\gamma_{-1}\gamma_{-1}^{\dagger}+\alpha_{1}^{2}\gamma_{-1}^{\dagger}\gamma_{-1}\gamma_{+1}\gamma_{+1}^{\dagger} + 
\alpha_{2}^{2}\gamma_{+2}^{\dagger}\gamma_{+2}\gamma_{-2}\gamma_{-2}^{\dagger}+\alpha_{2}^{2}\gamma_{-2}^{\dagger}\gamma_{-2}\gamma_{+2}\gamma_{+2}^{\dagger}+ \notag\right.\\
\left.+2\sum_{\beta}\sum_{\epsilon^{\prime}<-\Delta}(p_{\epsilon^{\prime}\beta}^{\star}p_{E_{\beta}\beta}h_{E_{\beta}\beta}^{\star}h_{\epsilon^{\prime}\beta}+h_{\epsilon^{\prime}\beta}^{\star}h_{E_{\beta}\beta}p_{E_{\beta}\beta}^{\star}p_{\epsilon^{\prime}\beta})(\gamma_{E_{\beta}\beta}^{\dagger}\gamma_{E_{\beta}\beta}-\gamma_{-E_{\beta}\beta}^{\dagger}\gamma_{-E_{\beta}\beta})\right).\label{eq:hint-final}
\end{gather}
Generally, the term in the second line cannot be expressed analytically. It, however, turns out to be irrelevant for the difference of transition frequencies for even and odd parities -- which is the quantity of interest -- as we now show. We begin by evaluating the frequency of the even-parity transition $\ket{0_1, 0_2} \leftrightarrow \ket{1_1, 1_2}$. To this end, we represent the many-body states in terms of eigenstate operators of the subgap states:
\begin{equation}
    \ket{0_1,0_2} = \gamma^\dagger_{-E_1 1} \gamma^\dagger_{-E_2 2} \ket{\Omega},\quad \ket{1_1,1_2} = \frac{1}{\sqrt{2}}\left(\gamma^\dagger_{E_1 1} \gamma^\dagger_{-E_1 1} - \gamma^\dagger_{E_2 2} \gamma^\dagger_{-E_2 2}\right) \ket{\Omega}.
\end{equation}
Here $\ket{\Omega}$ is a vacuum state for which $\gamma_{\pm E_1 1}\ket{\Omega} = 0$ and $\gamma_{\pm E_2 2}\ket{\Omega} = 0$. Note that the state $\ket{1_1,1_2}$ is a spin singlet state formed by quasiparticles in the two Andreev levels (within our model, the triplet states remain decoupled and can therefore be omitted).

To determine the correction to the transition frequency due to the interaction, we use first-order perturbation theory in $U_C$. We begin by finding the corrections to the energies of the states $\ket{0_1, 0_2}$ and $\ket{1_1, 1_2}$ due to $H_{\rm int}$ given by Eq.~\eqref{eq:hint-final}. This results in
\begin{gather}
    \delta E_{\ket{0_1, 0_2}} = U_C \left(\alpha_1^2 + \alpha_2^2 - 
    2\sum_{\beta}\sum_{\epsilon^{\prime}<-\Delta}(p_{\epsilon^{\prime}\beta}^{\star}p_{E_{\beta}\beta}h_{E_{\beta}\beta}^{\star}h_{\epsilon^{\prime}\beta}+h_{\epsilon^{\prime}\beta}^{\star}h_{E_{\beta}\beta}p_{E_{\beta}\beta}^{\star}p_{\epsilon^{\prime}\beta})
    \right),\quad \delta E_{\ket{1_1, 1_2}} = 0.
\end{gather}
The interaction thus shifts the frequency of the even transition by an amount
\begin{equation}
     {\hbar\delta \omega_\mathrm{even}} =  U_C \left(-\alpha_1^2 - \alpha_2^2 + 
    2\sum_{\beta}\sum_{\epsilon^{\prime}<-\Delta}(p_{\epsilon^{\prime}\beta}^{\star}p_{E_{\beta}\beta}h_{E_{\beta}\beta}^{\star}h_{\epsilon^{\prime}\beta}+h_{\epsilon^{\prime}\beta}^{\star}h_{E_{\beta}\beta}p_{E_{\beta}\beta}^{\star}p_{\epsilon^{\prime}\beta})\right).
\end{equation}
Discrete states in the odd parity sector are given by
\begin{equation}
    \ket{1_1,0_2} = \gamma^\dagger_{E_1 1} \gamma^\dagger_{-E_1 1} \gamma^\dagger_{-E_2 2} \ket{\Omega},\quad\ket{2_1,1_2} = \gamma^\dagger_{E_1 1} \gamma^\dagger_{E_2 2} \gamma^\dagger_{-E_2 2} \ket{\Omega}.
\end{equation}
Here, we focus on the spin up states -- in the absence of spin-orbit coupling the transition frequency is the same for the spin down states. The corrections to the energies of the discrete states due to the interaction read
\begin{gather}
    \delta E_{\ket{1_1, 0_2}} = U_C \left(\alpha_2^2 - 
    2\sum_{\epsilon^{\prime}<-\Delta}(p_{\epsilon^{\prime}2}^{\star}p_{E_{2}2}h_{E_{2}2}^{\star}h_{\epsilon^{\prime}2}+h_{\epsilon^{\prime}2}^{\star}h_{E_{2}2}p_{E_{2}2}^{\star}p_{\epsilon^{\prime}2})
    \right),\\ \delta E_{\ket{2_1, 1_2}} = U_C \left(\alpha_1^2 + 
    2\sum_{\epsilon^{\prime}<-\Delta}(p_{\epsilon^{\prime}1}^{\star}p_{E_{1}1}h_{E_{1}1}^{\star}h_{\epsilon^{\prime}1}+h_{\epsilon^{\prime}1}^{\star}h_{E_{1}1}p_{E_{1}1}^{\star}p_{\epsilon^{\prime}1})
    \right).
\end{gather}
Therefore, interaction shifts the frequency of the odd transition by an amount $\delta  \omega_{\rm odd}$ where
\begin{equation}
     {\hbar\delta \omega_\mathrm{odd}} = U_C \left(\alpha_1^2 - \alpha_2^2 + 2\sum_{\beta}\sum_{\epsilon^{\prime}<-\Delta}(p_{\epsilon^{\prime}\beta}^{\star}p_{E_{\beta}\beta}h_{E_{\beta}\beta}^{\star}h_{\epsilon^{\prime}\beta}+h_{\epsilon^{\prime}\beta}^{\star}h_{E_{\beta}\beta}p_{E_{\beta}\beta}^{\star}p_{\epsilon^{\prime}\beta}) \right).
\end{equation}
We thus find
\begin{equation}
     {\delta \omega_\mathrm{even} - \delta \omega_\mathrm{odd}} = \frac{U}{\hbar} = -2\alpha_1^2 \frac{U_C}{\hbar}.
\end{equation}
 {Substituting here the expression for $\alpha_1$ [cf.~Eq.~\eqref{eq:orthog}], we arrive to Eq.~\eqref{eq:delta-f}.}

\section{Solution for arbitrary tunneling matrix}\label{sec:non-diagonalt}
In the main text, when considering the microscopic model, we assumed that the first level on the dot was only connected to the first transport channel in both leads, and the second level on the dot was only connected to the second channel. Formally, this implied that the matrices describing the tunneling between the dot and the leads, $\hat{t}_L$ and $\hat{t}_R$ in Eq.~\eqref{eq:tun}, were diagonal. However, in general, these matrices do not have to be diagonal. Here, we consider the case of non-diagonal tunneling matrices and show that in the limit $\Gamma \gg \Delta${, all of our results remain the same (up to redefinitions of the parameters) as in the case of the diagonal matrices.} 
To illustrate this simplification, we determine the spectrum of the system for non-diagonal $\hat{t}_i$ and show that even in this case it is given by Eq.~\eqref{eq:energies} (with a proper redefinition of the channels). This observation is in line with general considerations for a junction with a short dwell time~\cite{Beenakker1991}.

Assuming that at $\varphi = 0$ the time-reversal symmetry is present in the system, the tunneling matrices $\hat{t}_i$ can be chosen to be real.  {Other than this constraint, $\hat{t}_i$ can have a general four-component form,}
\begin{equation}
    \hat{t}_i =
    \begin{pmatrix}
        t_{i,11} & t_{i,12}\\
        t_{i,21} & t_{i,22}
    \end{pmatrix}.
\end{equation}
Similarly to the diagonal case, the spectrum of the system is determined by the following Schr\"odinger equation
\begin{equation}\label{eq:big-schroed}
    E\left(1 + \frac{\sum_{i = R, L} \hat{\Gamma}_i}{\sqrt{\Delta^2 - E^2}}\right)C = \hat{\epsilon}_0\tau_z C + \frac{\Delta}{\sqrt{\Delta^2 - E^2}}
    \sum_{i = R, L}\begin{pmatrix}
    0 & \hat{\Gamma}_i e^{i\varphi_i}\\
    \hat{\Gamma}_i e^{-i\varphi_i} & 0 
    \end{pmatrix} C,
\end{equation}
where $\hat{\Gamma}_i = \pi \nu_0 \hat{t}_i \hat{t}_i^\dagger$ with $\nu_0$ being the normal-state density of states in the leads per spin projection. 
$\Gamma_{i,\alpha\beta}$ has a meaning of a normal-state tunneling rate of an electron from a dot level $\alpha$ to the $\beta$-th channel in lead $i = R$ or $L$. 
In the limit of strong tunneling, $\Gamma \gg \Delta$, equation~\eqref{eq:big-schroed} reduces to
\begin{equation}
    \label{eq:simple}
    E\sum_{i =  R, L} \hat{\Gamma}_i C = \Delta\sum_{i = R, L}
    \begin{pmatrix}
    0 &  \hat{\Gamma}_i e^{i\varphi_i}\\
    \hat{\Gamma}_i e^{-i\varphi_i} & 0 
    \end{pmatrix} C.
\end{equation}
Two positive definite matrices $\hat{\Gamma}_R$ and $\hat{\Gamma}_L$ can be simultaneously diagonalized by a phase-independent transformation. Application of this transformation in Eq.~\eqref{eq:simple} reduces the problem to that of the two uncoupled single-channel short junctions.
To construct the transformation, we first perform a unitary rotation that diagonalizes $\hat{\Gamma}_L$, $C = U_L C^\prime$. This changes $\hat{\Gamma}_L$ in Eq.~\eqref{eq:simple} to $U_L^\dagger \hat{\Gamma}_L U_L = \Lambda_L$, where $\Lambda_L$ is a diagonal matrix. Next, we stretch the spinor $C^\prime$ as $C^\prime = \Lambda^{-1/2}_L C^{\prime\prime}$. The two transformations map $\hat{\Gamma}_L$ to a unit matrix {[note that the latter of the two transformations is dimensionful; this is why the dimensionality changes]}. As a final step, we apply another unitary $U_R$ that diagonalizes $\Lambda_L^{-1/2} U_L^\dagger\hat{\Gamma}_R U_L \Lambda_L^{-1/2}$. If we now introduce a diagonal matrix $\lambda_R = U_R^\dagger \Lambda_L^{-1/2} U_L^\dagger\hat{\Gamma}_R U_L \Lambda_L^{-1/2} U_R$, we obtain the following equation:
\begin{equation}\label{eq:simplest}
    E C = \Delta
    \begin{pmatrix}
    0 &  \frac{e^{i\varphi_L} + \lambda_R e^{i\varphi_R}}{1 + \lambda_R}\\
    \frac{e^{-i\varphi_L} + \lambda_R e^{-i\varphi_R}}{1 + \lambda_R} & 0 
    \end{pmatrix} C
\end{equation}
Since Eq.~\eqref{eq:simplest} is diagonal in the channel space, it can be solved independently for the two channels. This yields two levels of the same form as in Eq.~\eqref{eq:energies}.

A chain of transformation similar to the one used when deriving Eq.~\eqref{eq:simplest} can be applied when calculating the matrix elements of the drive or the ionization rates. The outcome of this procedure is merely a redefinition of the dipole moment matrix $\hat{d}$ in Eq.~\eqref{eq:drive}. Therefore, for the purposes of the present manuscript, the general case with non-diagonal matrices $\hat{t}_i$ can be completely reduced to the diagonal case considered in the main text.

\section{AC-Stark shift}\label{sec:stark}
In establishing the dependence of the odd-state preparation rate $\gamma$ on the drive power $P$, we assumed that the drive frequency is at resonance with the transition frequency. However, there is an important caveat: the position of the resonance is itself sensitive to the power of the drive due to the ac-Stark shift~\cite{Townes1955}. 
 {Therefore, if one increases the power $P$ while keeping the drive frequency fixed, the drive would eventually go out of resonance with the transition.} Here, we establish the consequences of this effect for the high-power behavior of the odd-state preparation rate $\gamma$. 
 {Specifically, we demonstrate that for the fixed drive frequency, $\gamma$ saturates with the increase of $P$ at a much smaller value of $\gamma$ than that predicted by Eq.~\eqref{eq:sat_1}.}

To quantify the ac-Stark shift of the even-parity transition frequency, we note that the application of the drive effectively shifts the energy of the second Andreev level due to its coupling to the continuum. The shift of the second level, in turn changes the transition frequency from $\hbar\omega = E_1 + E_2$ to $\hbar\omega = E_1 + E_2 + \mathrm{Re}\,\Sigma_c(\omega)$. The real part of self-energy $\Sigma_c$ can be estimated from Eq.~\eqref{eq:sigmac},
\begin{equation}\label{eq:stark}
{\rm Re}\,\Sigma_c(\omega)={\rm Re}\,\sum_{c} \frac{\hbar|\Omega|^2|\alpha_c|^2}{2\omega + i0 - (E_1 + E_c)/\hbar } = - \frac{\left|\Omega\right|^2}{2\gamma_\mathrm{AC}},\quad \gamma_\mathrm{AC} = \frac{1}{2|\alpha_c|^2\nu_F\ln\frac{E_F}{\Delta}},
\end{equation}
where $E_F$ is the Fermi energy in the leads, $\nu_F$ is the normal-state density of states at the Fermi energy, and $\alpha_c$ is the dimensionless coupling at $E\sim E_F$.  {We note that this result is derived on the basis of the phenomenological Hamiltonian \eqref{eq:H}, which omits the off-resonant contributions in the drive term [see the discussion after Eq.~\eqref{eq:drive}]. In principle, such terms can produce a contribution to the ac-Stark shift of the same order as in Eq.~\eqref{eq:stark}, so the latter equation can only be viewed as an order of magnitude estimate.} Note that the argument of logarithm is large in practical conditions; this results in $\gamma_\mathrm{AC}\ll \gamma_{\rm s}$ [cf.~Eqs.~\eqref{eq:gamma_s_matrix_el} and \eqref{eq:stark}]. 

 {Equation~\eqref{eq:stark} implies that---even if the drive was resonant with the transition at small power---it becomes progressively more and more off-resonant as the power is increased [recall that $|\Omega|^2 \propto P$].  Using expression \eqref{eq:crossover_detuning} for the odd-state preparation rate in the case a drive detuned from the resonance, we find}
\begin{equation}
    \gamma = \frac{|\Omega|^2}{2\gamma_{\rm s}} \frac{|\Omega|^2}{|\Omega|^2 + \frac{\left|\Omega\right|^4}{4\gamma_\mathrm{AC}^2} + \kappa^2 / 8}.
\end{equation}
At high drive power, it saturates to
    \begin{equation}
        \gamma = \frac{2\gamma_\mathrm{AC}^2}{\gamma_{\rm s}}\ll \gamma_{\rm s}.
    \end{equation}
The saturation occurs at $|\Omega| \sim \gamma_{\rm AC} \ll \gamma_{\rm s}$.

 {Note that in the limit $\gamma_\mathrm{AC}\gg \gamma_{\rm s}$ (opposite to the one considered above) the saturation happens at $|\Omega|\sim\gamma_{\rm s}$, same as in the main text.}
\end{widetext}

\bibliography{references.bib}

\end{document}